\documentclass[qsecnum,amsmath,preprintnumbers,superscriptaddress,nofootinbib,aps,prd,10pt,a4paper]{revtex4-1}
\pdfoutput=1

\usepackage[utf8]{inputenc}
\usepackage{lipsum}
\usepackage{amsmath}
\usepackage{amsfonts}
\usepackage{amssymb}
\usepackage{amsthm}
\usepackage[colorlinks=black, citecolor=green, linkcolor=blue, linktocpage=true]{hyperref}
\usepackage{xcolor}
\usepackage{bm}
\usepackage{graphicx}
\usepackage{comment}
\usepackage{mathrsfs}
\usepackage{caption}
\usepackage{subcaption}
\usepackage{framed}
\usepackage{microtype}   
\usepackage{array}
\def\baselinestretch{1.2}
\begin{document}
\title{{\bf Deep Learning solutions to singular ordinary differential equations: from special functions to spherical accretion}}

\author{R. Cayuso}
\email[]{rcayuso@sissa.it}\thanks{Equal contribution.}
\address{SISSA. Via Bonomea 265, 34136 Trieste, Italy}
\address{IFPU - Institute for Fundamental Physics of the Universe, Via Beirut 2, 34014 Trieste, Italy}
\author{M. Herrero-Valea}
\email[]{mherrero@ifae.es}\thanks{Equal contribution.}
\address{Institut de Fisica d’Altes Energies (IFAE). The Barcelona Institute of Science and Technology. Campus UAB, 08193 Bellaterra (Barcelona) Spain}

\author{E. Barausse}
\email[]{barausse@sissa.it}

\address{SISSA. Via Bonomea 265, 34136 Trieste, Italy}
\address{IFPU - Institute for Fundamental Physics of the Universe, Via Beirut 2, 34014 Trieste, Italy}

\date{\today}

\begin{abstract}

Singular regular points often arise in differential equations describing physical phenomena such as fluid dynamics, electromagnetism, and gravitation. Traditional numerical techniques often fail or become unstable near these points, requiring the use of semi-analytical tools, such as series expansions and perturbative methods, in combination with numerical algorithms; or to invoke more sophisticated methods. In this work, we take an alternative route and leverage the power of machine learning to exploit Physics Informed Neural Networks (PINNs) as a modern approach to solving ordinary differential equations with singular points. PINNs utilize deep learning architectures to approximate solutions by embedding the differential equations into the loss function of the neural network. We discuss the advantages of PINNs in handling singularities, particularly their ability to bypass traditional grid-based methods and provide smooth approximations across irregular regions. Techniques for enhancing the accuracy of PINNs near singular points, such as adaptive loss weighting, are used in order to achieve high efficiency in the training of the network. We exemplify our results by studying four differential equations of interest in mathematics and gravitation -- the Legendre equation, the hypergeometric equation, the solution for black hole space-times in theories of Lorentz violating gravity, and the spherical accretion of a perfect fluid in a Schwarzschild geometry.

\end{abstract}

\maketitle
{\renewcommand{\baselinestretch}{1.2} \parskip=0pt
\setcounter{tocdepth}{2}
\tableofcontents}
\newpage

\section{Introduction}
Differential equations are ubiquitous in physics. They describe the rate of change of physical quantities under the action of external agents and provide the basic language in which physical laws are written. Unfortunately, the set of differential equations which allow for analytical solutions is small. Most equations cannot be solved in terms of elementary functions, nor can their solution be written in a closed form. Of course, this does not mean we cannot solve them. When small parameters are present, and if an analytic solution exists when those parameters vanish exactly, an approximate solution can be obtained using perturbation theory. Most modern particle physics is built using this method, with astounding success. When this approach is not feasible, numerical methods can be used to obtain solutions as a list of numerical values evaluated on a coordinate grid. These methods typically involve the discretization of the differential operators in the equation, and although convergence and robustness of this procedure can be subtle matters, depending on the problem at hand, their success is unquestionable. Numerical methods based on finite differences allow for the integration of complicated systems of non-linear coupled differential equations in partial derivatives, such as the Einstein or Navier-Stokes equations. 

A particular class of differential equations that can pose difficulties for
finite difference methods is provided by
those containing singular points. These appear when the coefficients multiplying the derivatives in the equation contain poles or other possible complex irregularities when approaching a point in the integration domain. In that case, solutions only exist in the neighborhood of the singular point if they behave in such a way as to cancel the irregular behavior, rendering the full equation (and solution) regular. This can be regarded as an extra condition for the solution, which replaces one of the boundary or initial conditions. As a consequence, not all possible boundary or initial data lead to solutions extending through the irregular points. In particular, numerical solutions starting from arbitrary boundary data will not lead, in general, to regular solutions. 

Equations with irregular points are common in physics, typically connected to critical behavior and to the existence of phase transitions in several systems, across diverse disciplines such as fluid mechanics, electromagnetism, quantum mechanics or gravitation. The standard way to solve these equations requires combining numerical and analytical methods.
For instance, one can obtain an asymptotic expansion of the {\it regular} solution around the irregular point, typically depending on some undetermined parameters. These can be fixed together with the boundary conditions through a shooting method, i.e., by integrating from the boundary and solving a set of algebraic equations demanding the matching of the solutions -- asymptotic and numerical --  close to the irregular point.
Alternatively, one can impose regularity of the solution at the irregular point by
including the latter in the computational grid,
and rewriting the equation at the irregular point using L'H\^{o}pital's rule. One can also expand the solution on a basis of functions regular at the irregular point, transforming the problem of solving the differential equation into that of finding the solution to an algebraic system.
Although possible, these approaches can be computationally heavy, as they require evaluating the equation several times, they are not guaranteed to converge for non-linear systems, and are tricky to implement in general (i.e., the implementation needs to be fine-tuned to the specific problem at hand).

In 2017, Raissi et al. \cite{raissi2017physics} introduced the concept of physics-informed neural networks (PINN), as a tool to solve differential equations by leveraging the efficiency of deep learning to fulfill the universal approximation theorem \cite{hornik1989multilayer}. A neural network (NN) is used to represent the (regular) solution of the equation, mapping points in the integration regime to the value of the solution. Their original motivation was data driven -- they proposed NNs as approximators of physical laws when solving unsupervised tasks which were constrained to also satisfy a set of equations -- but the presence of data to fit is not essential. In practice, PINNs work by substituting the loss function to be trained by back-propagation with a definite positive functional of the differential equation, bounded from below by its solution. By minimizing this functional, the NN learns the solution to the differential equation. For a detailed review, see \cite{hao2022physics}.

The main advantage of PINNs is that they are grid-independent. The solution is learned functionally, and it can be evaluated on points that were not in the original training set, while finite difference techniques require either running again the full integration for every different grid or interpolating. Although training can be computationally expensive, successive evaluations of a trained model are extremely cheap and can be performed very fast, with very few resources. Moreover, the development of deep learning methods has been exponential in the recent decade and will presumably continue in the next few years. 
Finally, it is worth mentioning that PINNs are also free from the convergence issues derived from the discretization of derivatives typically found when using finite differences. Therefore, bounds such as the Courant–Friedrichs–Lewy \cite{Courant2015OnTP} condition are irrelevant, since derivatives are computed by means of automatic differentiation, which provides their exact value, without resorting to discretization.

Focusing on the specifics of PINNs, one must note that as long as all layers and activation functions in the NN are analytical in the integration regime, its output will always be analytical as well. Hence, one can a priori argue that if a PINN is used to solve a differential equation with an irregular point, it will only be able to output the correct regular solution, with no need to impose regularity as a constraint. In this paper, we explore this possibility and show that PINNs can indeed find the correct regular solution with only an incomplete set of boundary data, as regularity through the singular points is already built in in the algorithm. 

Our work is organized as follows: In section \ref{sec:II}, we provide a brief review of differential equations with singular points. Section \ref{sec:III} introduces PINNs as a technique used to solve differential equations, while section \ref{sec:IIIa} details the network's architecture and training that we use in this work. In section \ref{sec:IV}, we perform experiments of solving equations with singular points by means of PINNs. The different examples are: The Legendre Differential Equation in section \ref{sec:IVa}, the Hypergeometric Equation in section \ref{sec:IVb}, the equations involved in solving for black holes in Lorentz violating gravity in section \ref{sec:aether}, and lastly the equations describing spherical accretion of a perfect fluid in a Schwarzschild geometry in section \ref{sec:accre}. Finally, in section \ref{sec:conclusions}
, we summarize our findings and provide concluding remarks, including potential future directions for research and applications of PINNs to problems with singular points. All computations in this work are done by using PyTorch \cite{NEURIPS2019_9015}.

\section{Differential equations with singular points}\label{sec:II}
In the following, we focus on systems of $N$ ordinary differential equations of the  form
\begin{align}\label{eq:system}
    \partial_x y_i(x) = f_i(x, y),\quad i=1,2,\dots N,\quad x\in \{x_0, x_1\}
\end{align}
where the functions $f_i(x,y)$ are generically non-linear, and $x\in \{x_0, x_1\}$, with $x_0<x_1$. Note that any higher-order ordinary differential equation can always be cast into this form by defining auxiliary variables. 

To obtain solutions, the system must be complemented with appropriate boundary data. For $N$ first-order equations, we need $N$ conditions, which we can take of the form
\begin{align}\label{eq:boundary}
    y_i(x_0)=z_i,\quad z_i \in \mathbb{R}.
\end{align}

The behavior of the system \eqref{eq:system} with boundary conditions \eqref{eq:boundary} will, however, depend on the behavior of the functions $f_i(x,y_i)$ over the integration domain. Let us exemplify this with the simplest case where all the functions are linear in the variables $y_i(x)$:
\begin{align}
    f_i(x,y) = \sum_{j=1}^N A^i_j(x) y_j.
\end{align}

In this case, we can classify the nature of the points $\hat x\in \{x_0,x_1\}$ according to the analytical behavior of the different $A_j^i(x)$, using Frobenius' method \cite{Frobenius1873} to obtain a power series expansion of the solution around any point. We can find three different types of behaviors \cite{tenenbaum1963ordinary}:
\begin{itemize}
    \item $\lim_{x\rightarrow \hat x}A_j^i(x) \in \mathbb{R}$ for all the functions $A_j^i(x)$. In this case, the point $\hat x$ is an ordinary point. Frobenius' method can be used to obtain a system of indicial equations by replacing $y_i \rightarrow (x-\hat x)^{r_i}$ in the differential equation, which has exactly $N$ independent solutions $r_i$.

    \item $\lim_{x\rightarrow \hat x}A_j^i(x)= \infty$ for some function, but $\lim_{x\rightarrow \hat x}(x-\hat x)^\alpha A_j^i(x) \in \mathbb{R}$ for some $\alpha>0$. In this case, at least one of the functions contains a pole of order $\alpha$ at the point $\hat x$. This is said to be a singular regular point. Frobenius's method still applies, and the system of inditial equations contains $N$ independent solutions.

    \item $\lim_{x\rightarrow \hat x}A_j^i(x)= \infty$ for some function, with the singularity different from a pole. In this case, we have an irregular point, and the space of solutions around it is harder to study in general. The existence of $N$ independent roots to the system of inditial equations is not guaranteed and depends on the Poincar\'{e} rank of the singularity. One must proceed by finding an asymptotic expansion of the solution around $\hat x$, and study its behavior independently.
\end{itemize}

A similar classification can be performed for non-linear functions $f_i(x,y)$, but the identification of singular points depends on implicit equations involving not only $x$ but also the solutions $y_i$. For instance, the equation 
\begin{align}
    (y-1)\frac{dy}{dx}+1=0,
\end{align}
will exhibit a singular point whenever $y=1$, but its position cannot be identified without solving the equation beforehand. By doing so, one finds that the singular point lies at $2\hat x = (-1 + y(0))^2$. For more complicated equations, a closed-form solution cannot be obtained. However,
one can still, for instance, look for an asymptotic solution near the (unknown) singular point, and apply a shooting method by using the singular point position as a shooting parameter. However, this approach can be tricky to apply in general, and one needs to proceed on a case-by-case basis. The equation in section~\ref{sec:aether} provides an example of this kind.

\section{Physics Informed Neural Networks}\label{sec:III}
PINNs are a machine learning technique used to solve problems involving differential equations, typically in the context of scientific studies of physical systems. They were first introduced in their current form in \cite{raissi2017physics} -- although previous works on the topic can be found, see e.g. \cite{lagaris1998artificial, doi:10.1137/140974596}  -- as a tool to fit experimental data constrained to obey specific physical laws. However, the existence of such data, although helpful for achieving large efficiency in specific problems, is not necessary. Instead, one can regard PINNs purely as numerical methods to solve differential equations. 

We start by introducing new variables $u_i^\theta(x)$, corresponding to representations of the variables $y_i(x)$ in \eqref{eq:system} that depend on a set of parameters $\theta$. Then, the problem of solving \eqref{eq:system} can be replaced by an optimization problem, corresponding to finding the values $\theta^*$ that minimize a loss function ${\cal L}$
\begin{align}\label{eq:theta_star}
  \theta^* =\underset{\theta}{\text{arg min}}\ {\cal L}= \underset{\theta}{\text{arg min}}\left\{ \sum_{i=1}^N \lambda_i \left\Vert\partial_x u_i^\theta(x)- f_i(x,u^\theta)\right\Vert^2\right\},
\end{align}
where all the $\lambda_i$ are positive non-vanishing constants. Since ${\cal L}$ is positive definite, its global minimum formally corresponds to the vanishing of all the terms in the sum, which then implies solving the system of differential equations. Boundary values can be implemented by adding terms to the loss functions of the form $\sim \left\Vert u_i^\theta(x_0)- z_i\right\Vert^2$ for all equations, but this can be impractical for ODEs\footnote{For PDEs instead, where boundaries are higher dimensional, this strategy can be more convenient.}. Instead, we impose them by changing variables and defining
\begin{align}\label{eq:redefinition}
    u_i(x)^\theta = z_i + g_i(x)n_i^\theta(x),
\end{align}
where $g_i(x)$ are monotonous functions on the integration regime, such that $g_i(x_0)=0$, and $n_i^\theta (x)$ is the output of a NN, which will be trained to minimize ${\cal L}$. This way, boundary conditions are automatically satisfied, provided that the NN representation $n^\theta (x)$ is smooth.

In practice, the $\mathbb{L}_2$ norm in \eqref{eq:theta_star} must be evaluated over a discretized set of $M$ collocation points $\hat{x}_a$ in the integration region, so that
\begin{align}
    \left\Vert\partial_x u_i^\theta(x)- f_i(x,u^\theta)\right\Vert^2 = \frac{1}{M}\sum_{a=1}^M\left(\partial_x u_i^\theta(\hat{x}_a)- f_i(\hat{x}_a,u^\theta)\right)^2.
\end{align}

The choice of these collocation points can influence the efficiency of the training of the NN used to solve the problem. In this work, we will always start with a regularly spaced grid, but we will allow it to change during the training procedure by resampling the collocation points randomly every certain number of epochs, as we will show later.

\subsection{Network architecture and training}\label{sec:IIIa}

\begin{figure}
    \centering
    \includegraphics[width=0.7\linewidth]{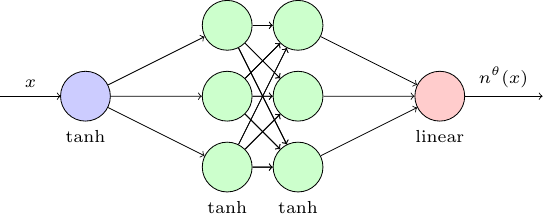}
    \caption{Example of a multilayer perceptron with two hidden layers, each consisting of three neurons.}
    \label{fig:ML-perceptron}
\end{figure}

We represent the outputs $n_i^\theta(x)$ by means of a set of $N$ feed-forward neural networks consisting each of a multilayer perceptron with $H_i$ hidden layers and $O_{ij}$ neurons in the layer $j$; and \emph{tanh} activation function in all layers\footnote{We also explored other activation functions -- $\sin$ and SiLU \cite{hendrycks2016gaussian} --with no appreciable differences.} but the last one, which has a linear output -- cf. figure \ref{fig:ML-perceptron} for a cartoon representation. The action of each layer is parametrized by a matrix $A_{ij}$ of dimension $O_{i(j-1)}\times O_{ij}$ and a vector $B_{ij}$ of length $O_{ij}$. The elements in $A_{ij}$ and $B_{ij}$ represent the trainable weights of the layer and are included in the full set $\theta$. If we label the output of layer $j$ in the network $i$ as $m_{ij}$, then
\begin{align}
    \nonumber &m_{i0}=m_{\rm input}=A_{i0} \cdot x +B_{i0},\\
    &m_{ij}= A_{ij}\tanh(m_{i (j-1)})+B_{ij}, \quad 1\leq i \leq H,
\end{align}
where both $A_{i0}$ and $B_{i0}$ are vectors of length $O_{i1}$. The final output of the network is hence
\begin{align}
    n_i^\theta(x) = A_{i\ {\rm output}}\cdot \tanh(m_{iH})+B_{i\ {\rm output}} = (m_{i\ {\rm output}}\circ m_{iH}\circ m_{i(H-1)}\dots m_{i1}\circ m_{i0})(x),
\end{align}
with the dimensions of $A_{i\ {\rm output}}$ and $B_{i\ {\rm output}}$ being $O_{iH}\times 1$ and $1$ respectively. Note that both operations, the action of the layer and the activation function $\tanh$ are smooth functions. Thus, the output $n^\theta(x)$ of the NN will always be smooth over the whole integration regime.

As described before, the objective loss function to minimize is given by the remainder of the differential equations evaluated over a set of collocation points
\begin{align}
    {\cal L}= \sum_{i=1}^N \lambda_i {\cal L}_i =  \frac{1}{M}\sum_{i=1}^N \sum_{a=1}^M \lambda_i \left(\partial_xu_i^\theta(\hat{x}_a)- f_i(\hat{x}_a,u^\theta)\right)^2,
\end{align}
which will, hence, depend on the output of the network and its derivatives. The latter will be computed without relying on any discretization by using automatic differentiation \cite{paszke2017automatic}.

We train the NN by using stochastic optimization in the form of the Adam algorithm \cite{kingma2014adam}. We start with a given learning rate $\eta_{\rm in}$, typically between $10^{-2}$ and $10^{-4}$, and we update it through the training by using a cosine annealing procedure with warm restarts \cite{loshchilov2016sgdr}. The learning rate at epoch $t$ will be thus given by
\begin{align}
    \eta (t) = \frac{\eta_{\rm in}}{2}\left(1+\cos \left(\frac{\hat{t}}{P}\pi\right)\right),
\end{align}
where $\hat t$ is the number of epochs since the last restart and $P$ is the period of the oscillation. When $\hat t=P$, a warm restart is produced -- $P$ is doubled, and $\hat t$ is set to zero again. This prevents the training from getting stuck at a local minimum.

Finally, in the case that $N>1$ and more than a single equation needs to be solved simultaneously, this becomes a multi-objective optimization problem, which can be difficult depending on the behavior of the individual terms in the loss function. In that case, we adopt an adaptive scheme to balance individual terms, given by \emph{SoftAdapt} \cite{heydari2019softadapt}. The parameters $\lambda_i$ are updated at each successive training epoch by following the rule
\begin{align}\label{eq:softadapt}
    \lambda_i = \frac{\exp \left[ T({\cal L}_i(t)-{\cal L}_i(t-1))\right]}{\sum_ i \exp \left[ T({\cal L}_i(t)-{\cal L}_i(t-1))\right]},
\end{align}
where ${\cal L}_i(t)$ refers to the loss term corresponding to equation $i$ evaluated at epoch $t$, and $T$ is a temperature factor that weights how much the differences in loss values translate to changes in $\lambda_i$.

Finally, in order to enforce the network to learn the functional form of the solution, instead of simply over-fitting its value over the collocation points, we randomly resample those every $S$ training epochs, by
\begin{align}
    \hat x_a = x_1 + (x_0 - x_1) \hat P_a,
\end{align}
where $\hat P_a$ is a point sampled from a uniform distribution in the interval $[0,1)$.

\section{Experiments}\label{sec:IV}

\subsection{Legendre Differential Equation}\label{sec:IVa}

Let us start by discussing a simple differential equation of interest in mathematics displaying singular regular points -- the Legendre equation
\begin{align}\label{eq:Legendre}
    (1-x^2)\frac{d^2y}{dx^2}-2x \frac{dy}{dx}+l(l+1)y = 0,
\end{align}
where $l\in \mathbb{R}$. This is a second-order differential equation, so it will, in general, have two linearly independent solutions. Dividing by $(1-x^2)$, we also observe that it contains two regular singular points due to poles at $x=\pm 1$. At these points, the pair of solutions have different behaviors -- one is smooth, corresponding to a Legendre function of the first kind, and one diverges, a Legendre function of the second kind. For integer $l$, the smooth solution corresponds to the Legendre polynomial of order $l$.

If we were to solve this equation by standard numerical methods based on finite differences,
a shooting algorithm could be used. By imposing regularity at both $x=\pm 1$, the required number of boundary conditions is already fulfilled. However, integration by finite differences would also require the value of the derivative of the solution in order to start integrating from one of the boundaries. The way to proceed is to leave this derivative as a free parameter, integrate from both boundaries starting from an asymptotic solution, and solve a system of equations demanding the matching of the solution and its first derivative at some intermediate point. Of course, every step in the root-finding algorithm used to solve this would require a full integration of the differential equation, 
making this approach rather inefficient.

Instead, PINNs can solve this problem without requiring any semi-analytical expansion nor any root-finding algorithm. Moreover, since the output of our NN is necessarily smooth by construction, it will always capture the regular solution, even if we only give it the boundary condition at one of the boundaries. The requirement of smoothness in the second boundary, which is here implicit by the procedure, can be regarded as the second condition. We thus implement only the boundary condition at $x=-1$ through a redefinition of the network variable of the form \eqref{eq:redefinition}. For integer $l$,  this leads to
\begin{align}
    y^\theta(x) = (-1)^l + \left(1-e^{-(x+1)}\right)n^\theta(x),
\end{align}
where we have chosen $g(x) = 1-e^{-(x+1)}$. The value of the boundary condition is arbitrary, since the equation is invariant under constant rescalings, but the sign matters for the orthogonality of the solutions. Here, we have adopted the same convention as the normalized Legendre polynomials, in order to draw comparisons easily.

\begin{figure}[]
    \centering
    \begin{subfigure}[b]{0.45\textwidth}
        \centering
        \includegraphics[width=\textwidth]{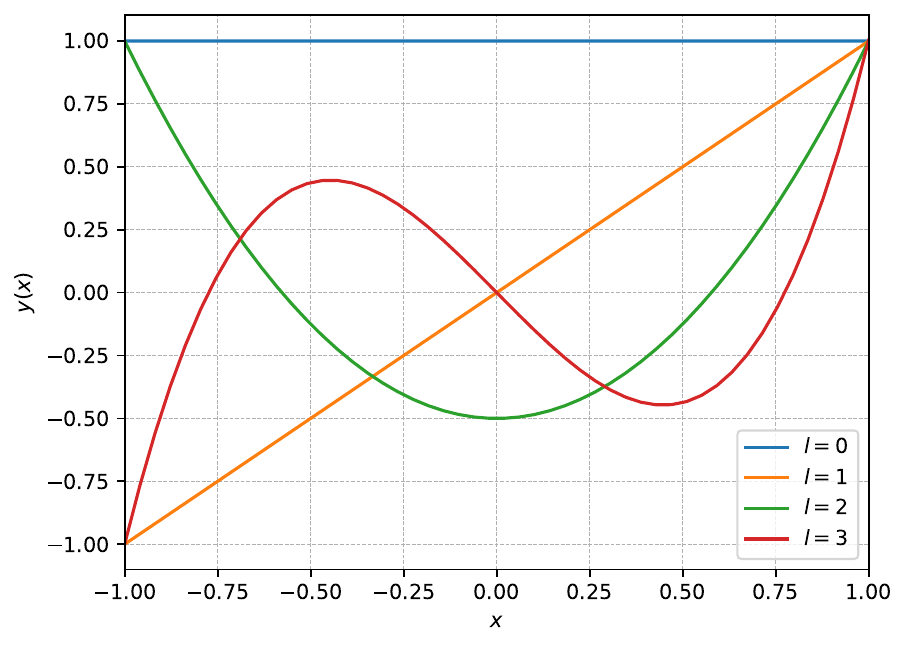}
        \caption{First Legendre polynomials, obtained using PINNs.}
    \end{subfigure}
    \hfill
    \begin{subfigure}[b]{0.45\textwidth}
        \centering
        \includegraphics[width=\textwidth]{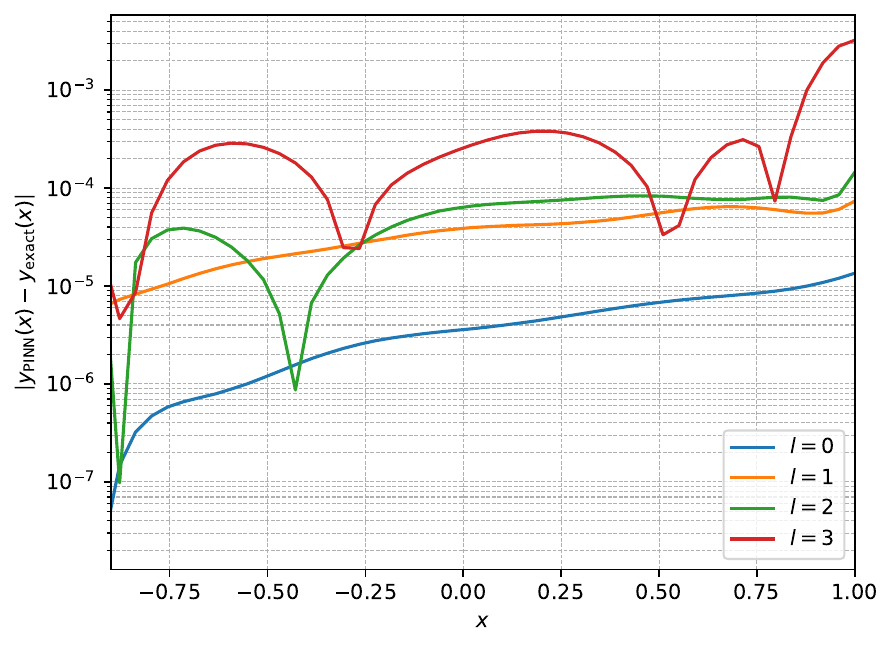}
        \caption{Residuals for the PINN solutions, compared to analytical results.}
    \end{subfigure}
    \caption{Benchmark results for the solution of Legendre differential equation \eqref{eq:Legendre} for $l=\{0,1,2,3\}$. All solutions were obtained with a NN with $H=4$ hidden layers and $O=32$ neurons in each layer. We also used the same number of training epochs for all solutions.}
    \label{fig:Legendre_pol}
\end{figure}


Finally, let us highlight that we refrain from rewriting the equation as a first-order system for this case and instead retain the form \eqref{eq:Legendre} to avoid dealing with a multi-objective optimization problem. Thus, our loss function is simply
\begin{align}
    {\cal L}_{\rm Legendre}= \frac{1}{M}\sum_{a=1}^M\left((1-\hat x_a^2)\left.\frac{d^2y^\theta}{dx^2}\right|_{\hat x_a}-2\hat x_a \left.\frac{dy^\theta}{dx}\right|_{\hat x_a}+l(l+1)y^\theta\right)^2,
\end{align}
and there is no need to use adaptative learning. 

\begin{figure}[]
    \centering
    \begin{subfigure}[b]{0.45\textwidth}
        \centering
        \includegraphics[width=\textwidth]{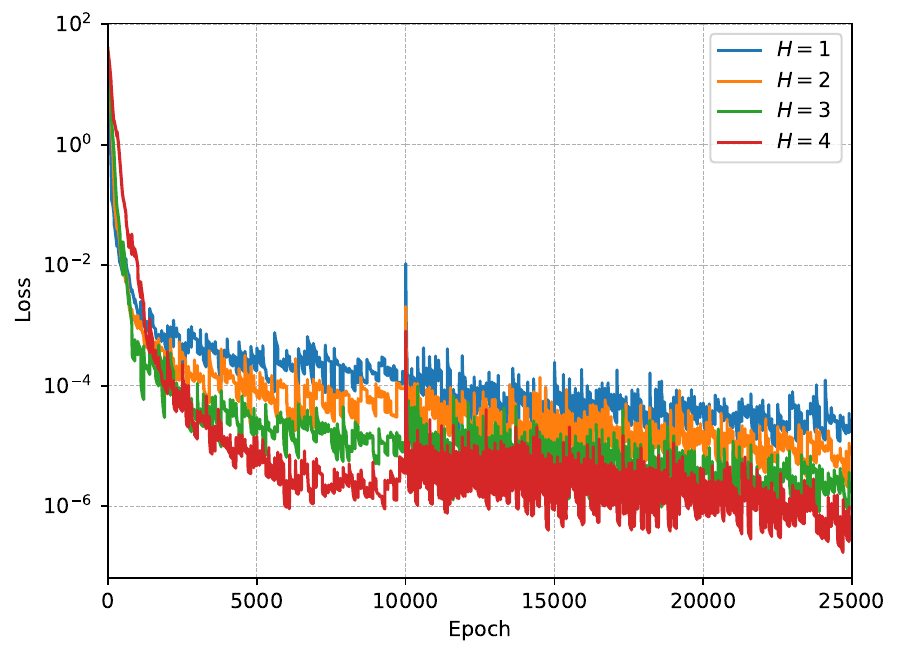}
        \caption{Evolution of the loss function for different number of hidden layers $H$, all of them with $O=64$ neurons.}
    \end{subfigure}
    \hfill
    \begin{subfigure}[b]{0.45\textwidth}
        \centering
        \includegraphics[width=\textwidth]{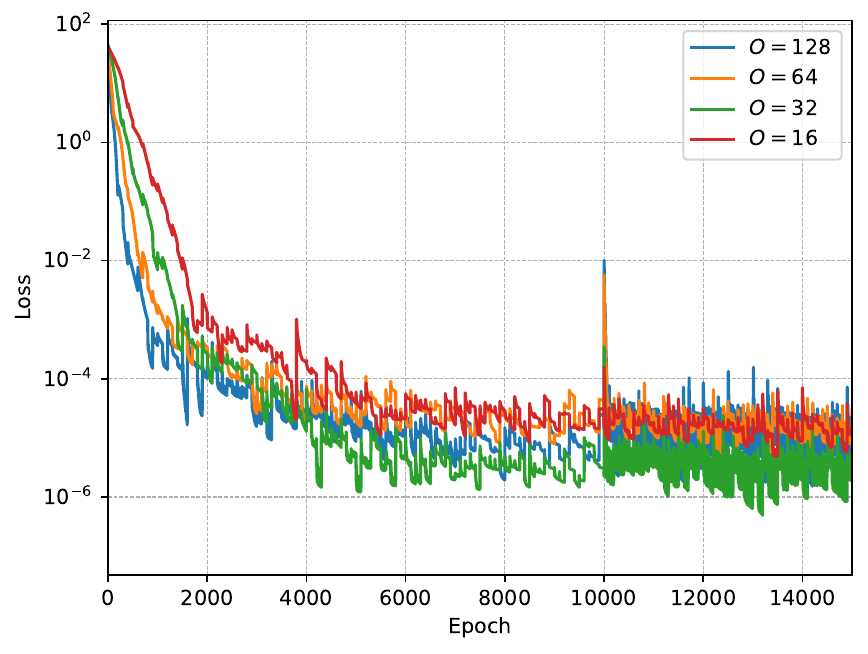}
        \caption{Evolution of the loss function for different number of neurons in the hidden layers. Here, we kept their number fixed to $H=3$.}
    \end{subfigure}
    \caption{Loss function as a function of the number of epochs, when solving the Legendre differential equation for $l=2$ with different number of layers and neurons in the NN. All experiments are performed using the hyperparameters in table \ref{tab:legendre}. The spikes in the plots correspond to the warm restarts of the cosine annealing scheduler.}
    \label{fig:Legendre_layers}
\end{figure}

We performed several experiments to test the robustness of our approach. The hyperparameters used are summarized in table \ref{tab:legendre}. Benchmark results for the first few Legendre polynomials are shown in figure \ref{fig:Legendre_pol}. We observe that at a fixed size of the network and training epochs, the result of the PINN is worse as the solution becomes more complex. We can test the origin of this dependence by increasing the number of hidden layers and neurons -- cf. figure \ref{fig:Legendre_layers}. We observe that while a large number of layers yields a substantial reduction of the loss value, there is not a clear gain by increasing the number of neurons beyond $\sim 32$ neurons per layer, where it reaches saturation. All experiments were performed with $M=50$ collocation points; however, we observe that such saturation is only reached at a higher number of neurons per layer if the number of collocation points is increased. Fluctuations in the value of the loss correspond to the resampling of the collocation points, which nonetheless allows for an improvement of the training by decreasing its trend value.

\begin{table}
    \centering
    \begin{tabular}{|c||c|c|c|c|c|}
        \hline
        Hyperparameter & $\eta_{\rm in}$ & Adam $\beta$'s & $P$ & $S$ & $M$ \\
        \hline
        Value& $10^{-3}$ & (0.9,0.9) &  $10^4$ & $10^2$ & 50 \\
        \hline
    \end{tabular}
    \caption{Hyperparameters chosen for solving the Legendre differential equation.}
    \label{tab:legendre}
\end{table}

\subsection{Hypergeometric Equation}\label{sec:IVb}
We focus now on a second example of a differential equation with singular points on the real axis -- the hypergeometric equation
\begin{align}\label{eq:hyper}
    x(x-1)\frac{d^2y}{dx^2}+\left(c-x(a+b+1)\right)\frac{dy}{dx}-ab y =0,
\end{align}
where $a,b$ and $c>0$ are real coefficients. This equation has two singular points for finite $x$, corresponding to $x=0$ and $x=1$, and a third one laying at $x\rightarrow \infty$. All are singular regular points as per the classification in section \ref{sec:II}. Any second-order differential equation with three regular singular points can be rewritten as \eqref{eq:hyper} by a change of variables. Note also that the equation is symmetric with respect to $a\leftrightarrow b$.

Solutions to this equation are given in terms of elementary functions -- powers and logarithms -- multiplying the hypergeometric series
\begin{align}
    {}_2 F_1(a,b,c;x) = \sum_{n=0}^\infty \frac{(a)_n (b)_n}{(c)_n}\frac{x^n}{n!},
\end{align}
where $(z)_n$ is the Pochhammer symbol
\begin{align}
    (z)_n=\begin{cases}
        1,&z=0\\
        z(z+1)\dots (z+n-1), &z>0
    \end{cases}.
\end{align}

\begin{table}
    \centering
    \begin{tabular}{|c||c|c|c|c|c|}
        \hline
        Hyperparameter & $\eta_{\rm in}$ & Adam $\beta$'s & $P$ & $S$ & $M$ \\
        \hline
        Value& $3\times 10^{-4}$ & (0.9,0.9) &  $8\times 10^3$ & $10^2$ & $10^2$ \\
        \hline
    \end{tabular}
    \caption{Hyperparameters chosen for solving the hypergeometric equation.}
    \label{tab:hyper}
\end{table}

For positive $a,b$, the series ${}_2 F_1(a,b,c;x)$ has a finite radius of convergence $|x|<1$, with the function diverging at the singular point. However, if either $a$ or $b$ are negative integers, the series is truncated, and the hypergeometric function reduces to a polynomial
\begin{align}
    {}_2 F_1(-m,b,c;x) = \sum_{n=0}^m (-1)^n\binom{m}{n}\frac{(b)_n}{(c)_n}z^n,
\end{align}
which is regular everywhere and solves the equation \eqref{eq:hyper}.

We again resort to PINNs to prove their efficiency in solving \eqref{eq:hyper} through its singular points. Since we look for regular solutions, we fix $a=-m$ and vary all three coefficients to test different cases. At $x=0$, the zeroth term in the series provides the boundary condition ${}_2 F_1(a,b,c;0)=1$, and hence we parametrize our solution in terms of a NN by
\begin{align}
    y^\theta(x)= 1 + x n^\theta(x),
\end{align}
where we have chosen in this case $g(x)=x$. Note, however, that the value of the solution at the second singular point $x=1$ is not fixed and must be learned by the NN, which must then oppose the divergence of the terms in the equation.

To construct the loss functions, we again refrain from rewriting this equation in terms of a first-order system to avoid a multi-objective problem. Hence, we will use
\begin{align}
    {\cal L}_{\rm Hyper}= \frac{1}{M}\sum_{a=1}^M\left(\hat x_a(\hat x_a-1)\left.\frac{d^2y^\theta}{dx^2}\right|_{\hat x_a}+\left(c-\hat x_a(a+b+1)\right)\left.\frac{dy^\theta}{dx}\right|_{\hat x_a}-ab y^\theta\right)^2,
\end{align}

\begin{figure}[]
    \centering
    \begin{subfigure}[b]{0.45\textwidth}
        \centering
        \includegraphics[width=\textwidth]{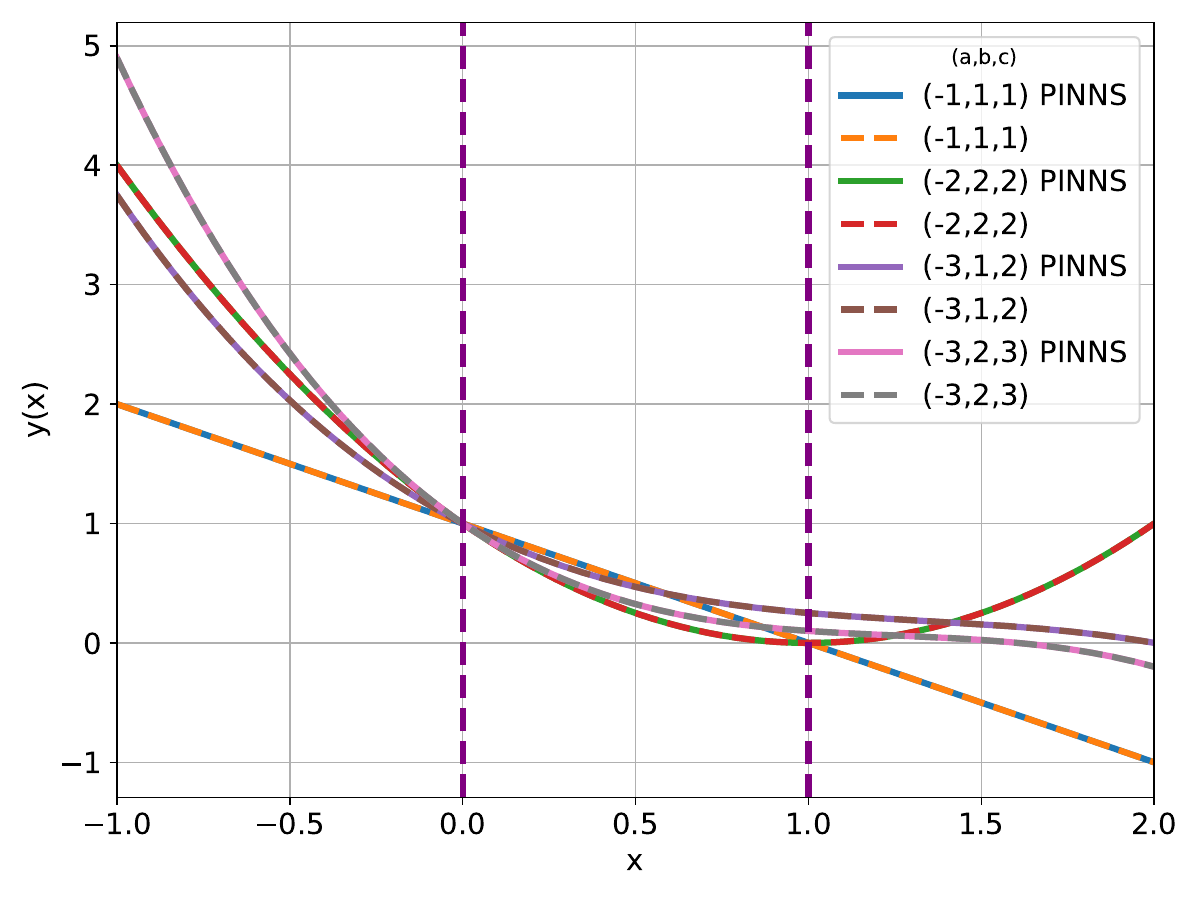}
        \caption{Some hypergeometric functions, obtained using PINNs contrasted to the analytical solutions.}
    \end{subfigure}
    \hfill
    \begin{subfigure}[b]{0.45\textwidth}
        \centering
        \includegraphics[width=\textwidth]{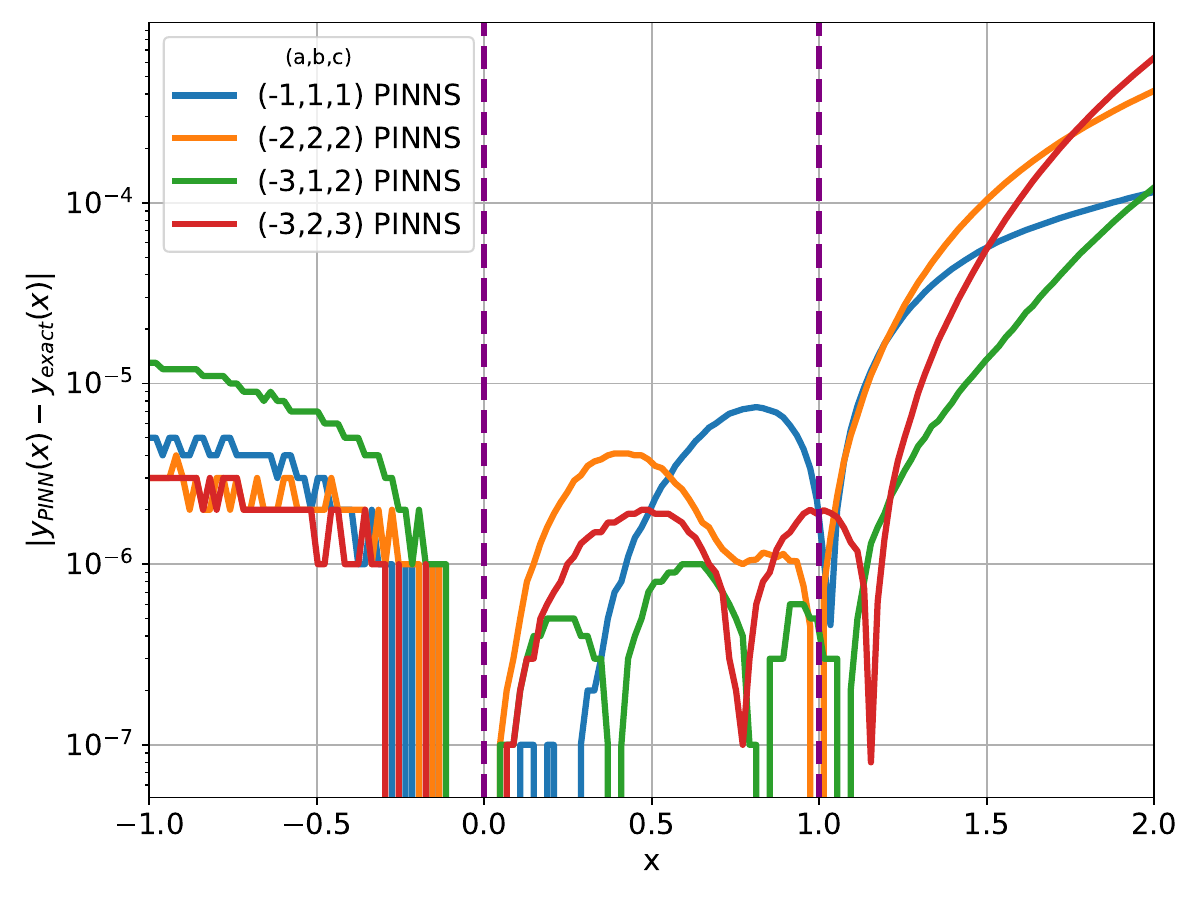}        
        \caption{Residuals for the hypergeometric solutions, compared to analytical results.}
    \end{subfigure}
    \caption{Benchmark results for the hypergeometric equation \eqref{eq:hyper}. All solutions were obtained with a NN with a single $H=2$ hidden layer and $O=64$ neurons. The vertical dashed lines in magenta denote the positions of the singular points $x=0$ and $x=1$.}
    \label{fig:hyper}
\end{figure}

\begin{figure}[]
    \centering
    \begin{subfigure}[b]{0.45\textwidth}
        \centering
        \includegraphics[width=\textwidth]{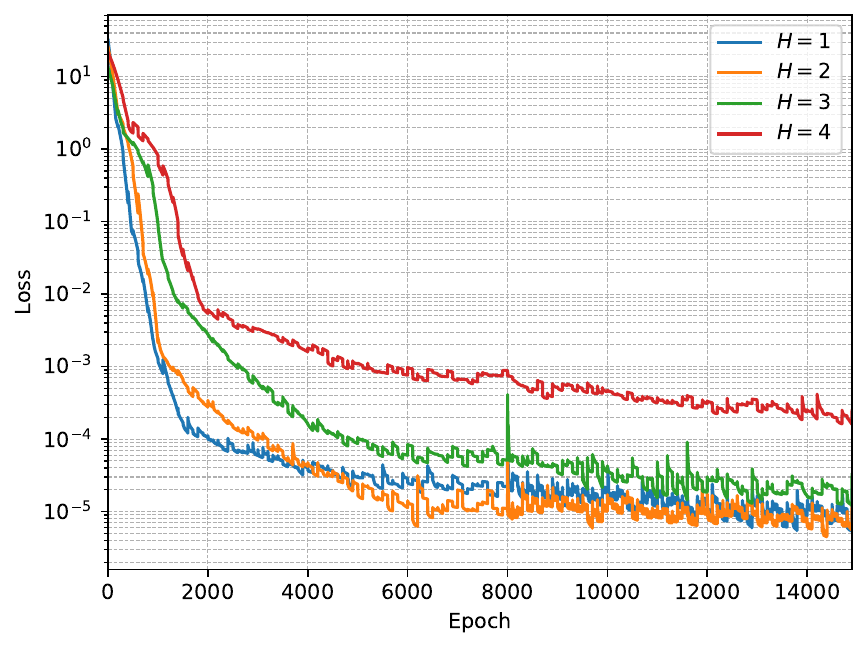}
        \caption{Evolution of the loss function for different number of hidden layers $H$, keeping $O=32$ neurons in each layer.}
    \end{subfigure}
    \hfill
    \begin{subfigure}[b]{0.45\textwidth}
        \centering
        \includegraphics[width=\textwidth]{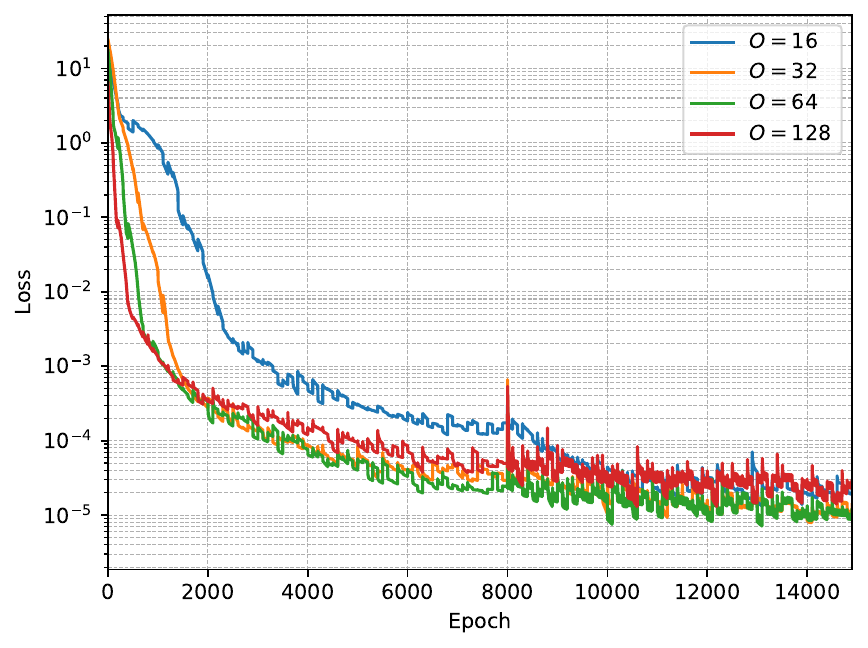}
        \caption{Evolution of the loss function for different number of neurons in a model with a single $H=1$ hidden layer.}
    \end{subfigure}
    \caption{Loss function in terms of the number of epochs for solving the hypergeometric equation for $(a,b,c)=(-2,2,2)$ in the domain $x=\{x=0,1.2\}$ with different number of layers and neurons in the NN.}
    \label{fig:hyper_layers}
\end{figure}

The set of hyperparameters chosen for these experiments is shown in table \ref{tab:hyper}, while benchmark results for a bunch of solutions can be found in figure \ref{fig:hyper}. Note that all of them are perfectly smooth through the singular points at $x=0$ and $x=1$. The error of the PINNs solutions tends to grow past the $x=1$  point, we noted that to keep this error under control it was necessary to train the network by employing domain decomposition. If the total spatial domain is given by $\mathcal{D}=\{ x_{b}, x_{f}\} $, we first train the network in a domain $\mathcal{D}_{i} = \{x_{b},x_{i}\}$, where $x_{i} < x_{f}$, until the loss is below a desired threshold $\mathcal{T}_{\mathcal{L}}$. Subsequently, the training continues in a domain  $\mathcal{D}_{i+1} = \{x_{b},x_{i+1}\}$, where $x_{i} < x_{i+1}  <x_{f}$. This procedure can be iterated until reaching the desired spatial domain $\mathcal{D}$. Note that here, unlike in the case of the Legendre equation \eqref{eq:Legendre}, we find it more efficient to work with shallow networks but a larger number of neurons -- see figure \ref{fig:hyper_layers}.

\subsection{Black holes in Lorentz violating gravity}\label{sec:aether}

Lorentz violations in gravity have received considerable attention in the last fifteen years \cite{MHV-review-HG}, as they allow for building a theory of gravity that is renormalizable at power-counting level, and potentially also at perturbative level~\cite{Jacobson:2000xp,Horava:2009uw,Blas:2009qj,Blas:2010hb,Barvinsky:2015kil,Barvinsky:2019rwn,Bellorin:2022qeu}. The concept of a black hole in these theories, however, is problematic to define~\cite{Blas:2011ni,Barausse:2011pu}, because Lorentz violations generally allow for multiple propagation speeds for the various gravitational degrees of freedom. Moreover, these speeds are superluminal~\cite{Elliott:2005va} and even diverging at high energies. This has prompted several investigations of the causal structure of black holes in these theories, both in spherical symmetry~\cite{Eling:2006ec,Blas:2011ni,Barausse:2011pu,Barausse:2013nwa,Lara:2021jul} and beyond~\cite{Barausse:2015frm,Adam:2021vsk}. A common feature of the solutions that have been found is the presence of multiple event horizons -- for the various degrees of freedom present in the theory -- and, at least in spherical symmetry, the presence of a universal horizon, i.e., quite surprisingly, a causal boundary for signals of arbitrarily large speed~\cite{Blas:2011ni,Barausse:2011pu}. 

Understanding the structure and properties of the multiple horizons of black holes in Lorentz-violating gravity is not just a mathematical curiosity, but
is also crucial to understand whether black hole thermodynamics holds in these theories~\cite{Berglund:2012bu,Cropp:2013sea,DelPorro:2022kkh,DelPorro:2023lbv,Herrero-Valea:2020fqa}, and to understand their viability in the light of astrophysical black hole observations~\cite{Ramos:2018oku,Franchini:2021bpt}. Most properties of standard event horizons in general relativistic models must be translated to the universal horizon for the theory to be sensible against our accumulated understanding of gravitational phenomena. However, solving for black hole geometries in Lorentz violating gravity is numerically involved, as the equations present singular points at the locations of the various horizons~\cite{Eling:2006ec,Blas:2011ni,Barausse:2011pu,Barausse:2015frm}. In fact, besides solutions that are regular at these horizons, unphysical solutions that are singular there also exist.

To illustrate these intricacies in a simple case, we will now consider the structure equations for spherical static black holes in the decoupling limit, i.e. the limit in which the Lorentz-violating fields couple weakly to the standard tensor gravitons of general relativity \cite{Blas:2011ni}. The metric is, therefore, provided by the usual Schwarzschild metric of general relativity, and one has to solve only for the Lorentz-violating fields. Moreover, because of the spherical symmetry, only the scalar Lorentz violating degree of freedom is non-trivial. In more detail, the dynamics of this scalar graviton is described by a single ordinary differential equation:
\begin{equation}\label{eq:aether}
\begin{split}
A(x) \left[\left(s_0^2-1\right) \left(-(x-1)^2\right) A(x)^4+2 \left(s_0^2+1\right)
   (x-1) A(x)^2-s_0^2+1\right] A''(x)\\-\frac{2 A(x)^2 \left[\left(s_0^2+1\right) (x-1)
   x^2 A'(x)^2+s_0^2\right]}{x^2}-2 \left(s_0^2-1\right) (x-1) A(x)^5 A'(x)\\+2
   \left(s_0^2+1\right) A(x)^3 A'(x)+2 \left(s_0^2-1\right) A'(x)^2+2 s_0^2
   \left(\frac{1}{x}-1\right)^2 A(x)^6=0
\end{split}
\end{equation}
where $x = 2M/r$ ($r$ being the radial Schwarzschild coordinate and $M$ the black hole mass), $A(x)$ is the field, $s_{0}$ is the spin-0 mode propagation speed, and primes denote derivatives with respect to $x$.  The notation employed here is the same as in Refs.~\cite{Barausse:2011pu,Barausse:2013nwa}, i.e. $A$ is the \ae ther component $u^v$ in Eddington-Finkelstein coordinates. At spatial infinity $A(x\to 0) =1 $ in order to comply with asymptotic flatness. However, not only is this boundary condition tricky to impose because the equation has a singular point at $x=0$, but an additional singular point appears at the spin-0 (i.e. scalar) horizon $x=x_s$, where the coefficient of the highest derivative $A''(x)$ vanishes
\begin{equation}\label{regspin0}
    \left(s_0^2-1\right) \left(-(x_s-1)^2\right) A(x_s)^4+2 \left(s_0^2+1\right)
   (x_s-1) A(x_s)^2-s_0^2+1=0\,.
\end{equation}

This equation thus introduces an extra level of complexity with respect to the equations discussed in the previous sections, where the position of the singular points was explicit. To find physical solutions with standard methods, one then has to impose regularity at $x=0$ and at the spin-0 horizon. The former can be imposed by Taylor expanding the equation, which yields the regular solution $A(x)=1+x/2+a_2 x^2+O(x)^3$, where $a_2$ is an undetermined integration constant. The latter is also imposed by assuming a Taylor expansion $A(x)=A_0+A_1 (x-x_s)+ A_2 (x-x_s)^2+O(x-x_s)^3$ around the spin-0 horizon, with $A_0$ and $x_s$ related by~\eqref{regspin0}. Note, however, that this equation is not linear, and one therefore needs to check which solution branch yields a physical solution. Similarly, the coefficients $A_1$ and $A_2$ are determined once $A_0$ and $x_s$ are fixed, but only up to the choice of
the solution branch -- since $A_1$ is determined by a quadratic equation.
Modulo this ambiguity, one can then perform two numerical integrations -- one
from $x=\epsilon$  and one from $x=x_s-\epsilon$, with $\epsilon$ a small number -- with initial conditions provided by the two Taylor expansions, and match the two solutions by imposing continuity of $A$ and $A'$ at an intermediate point by choosing $x_s$ and $a_2$ (bilateral shooting). This can be tricky as the convergence of the procedure depends
critically on the initial guess of $x_s$ and $a_2$.

In contrast to that, finding solutions for equation \eqref{eq:aether} by means of PINNs requires no regularity conditions and no matching of solutions at all. As done in the previous sections, we only need to define a parametrization for the $A(x)$ field that adequately imposes the boundary condition at one of the edges of the integration regime $A(0)=1$,
\begin{equation}\label{eq:param_aether}
    A^{\theta}(x) = A(0)e^{xn^{\theta}(x)},
\end{equation}
and an adequate definition for the loss function,
\begin{equation}
    \mathcal{L}_{\ae ther} = \frac{1}{N}\sum_{a=1}^N\left( \mathcal{E}_{\ae ther}(A^\theta)  \right)^{2} \left(\frac{ \hat{x}_{a}^{2}}{A^{\theta}(\hat{x}_{a})^{3}}\right)^2, 
\end{equation}
where $\mathcal{E}_{\ae ther}$ is the left-hand-side of equation \eqref{eq:aether} evaluated on the NN representation of $A(x)$, and we have introduced a regulator term which does not modify the position of the global minimum of the loss function, but helps in enforcing the solution not to collapse onto a vanishing function -- which is also solution to \eqref{eq:aether} but does not agree with the boundary conditions --, and in giving more weight to the region of large $x$ (small radius) where convergence is more difficult. Let us note that the choice for the parametrization of the solution in \eqref{eq:param_aether} differs slightly from that introduced in \eqref{eq:redefinition}, as here we have used an exponential representation in order to enforce the solution to be positive definite. 

\begin{figure}[]
    \centering
    \begin{subfigure}[b]{0.45\textwidth}
        \centering
        \includegraphics[width=\textwidth]{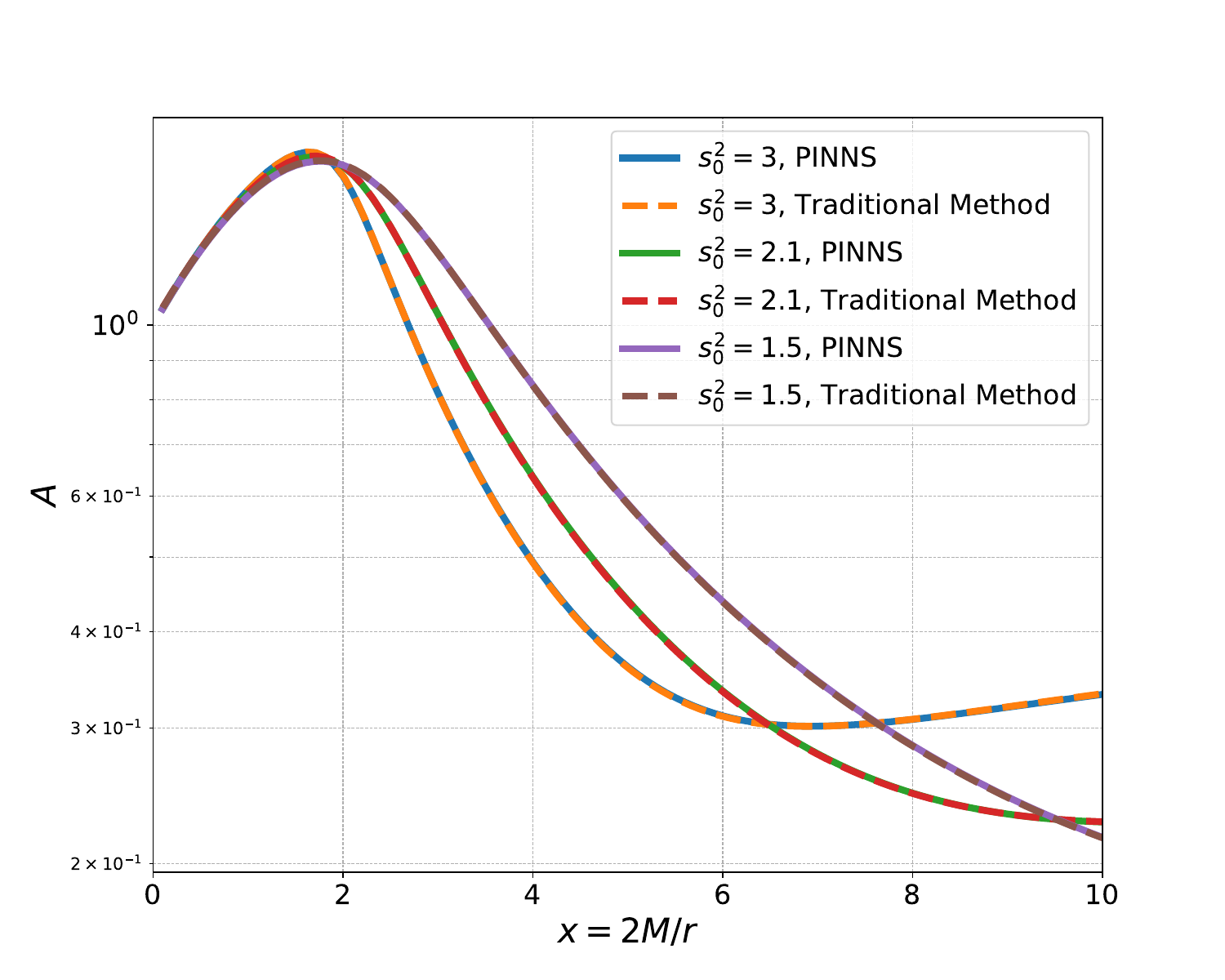}
        \caption{Solutions for the $A(x)$ field for different values of the spin-0 mode propagation squared speed $s_{0}^2$.}
        \label{fig:aetherfield}    
    \end{subfigure}
    \hfill
    \begin{subfigure}[b]{0.45\textwidth}
        \centering
        \includegraphics[width=\textwidth]{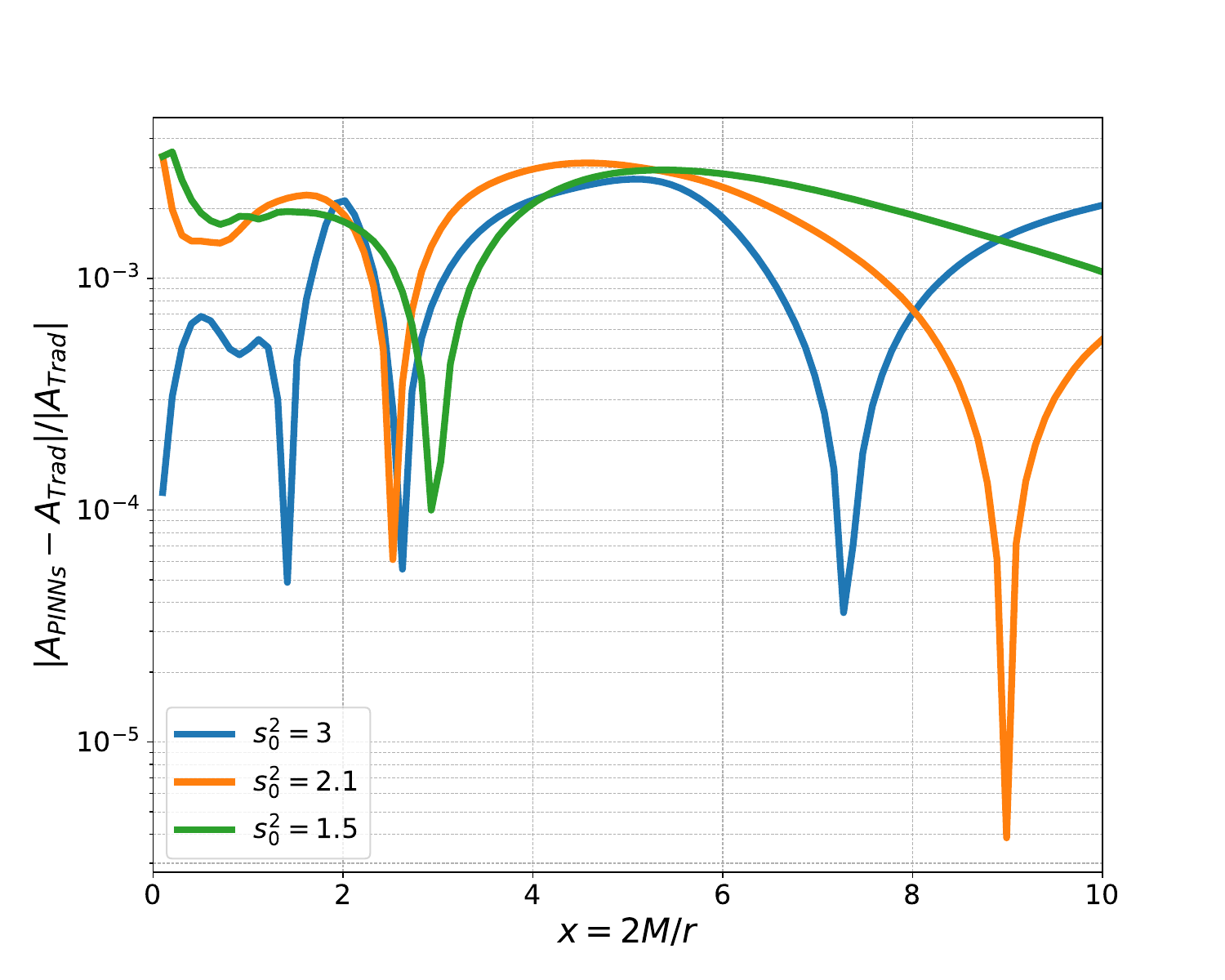}
        \caption{Relative error between the PINNs solutions and the traditional solutions for the $A(x)$ field.}
        \label{fig:aethererror}
    \end{subfigure}
    \caption{Results for the  Lorentz violating 
    scalar graviton $A$,    
    for different values of the spin-0 propagation speed squared $s_{0}^{2}$, obtained using PINNs, compared to traditional methods. All PINN solutions were obtained with a NN with $H=6$ hidden layers and $O=32$ neurons in each layer. For reference, these solutions are found for $M=0.5$, for which, the spin-0 horizon lies at $x_{s}= 1.11$ for $s_{0}^{2}=3$, $x_{s}= 1.08$ for $s_{0}^{2}=2.1$, and $x_{s}= 1.05$ for $s_{0}^{2}=1.5$.}
    \label{fig:aethersols}
\end{figure}

\begin{figure}[]
    \centering
    \begin{subfigure}[b]{0.45\textwidth}
        \centering
        \includegraphics[width=\textwidth]{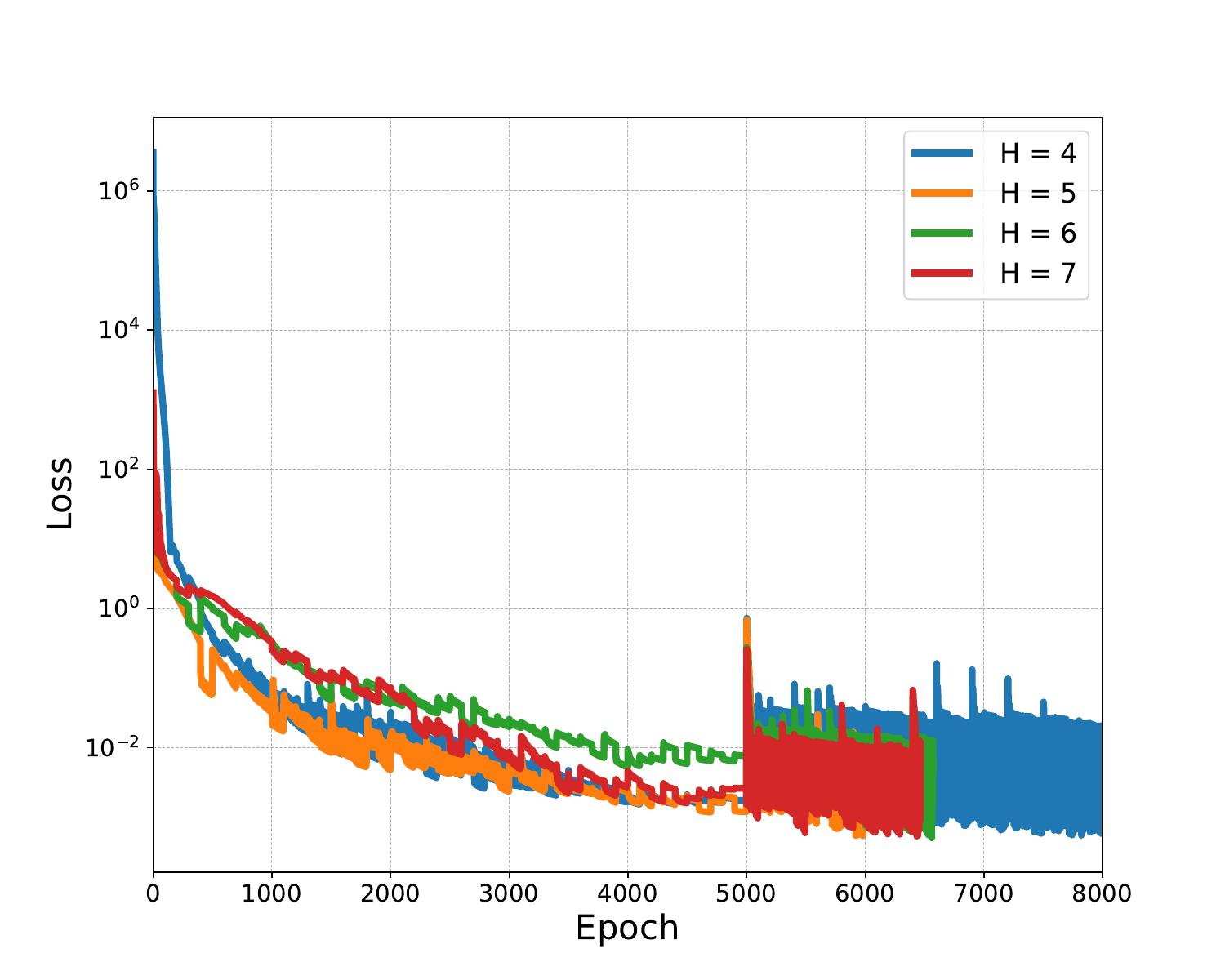}
        \caption{Evolution of the loss function for different number of hidden layers H, keeping $O=128$ neurons in each layer.}
        \label{fig:aether_loss_layers}
    \end{subfigure}
    \hfill
    \begin{subfigure}[b]{0.45\textwidth}
        \centering
        \includegraphics[width=\textwidth]{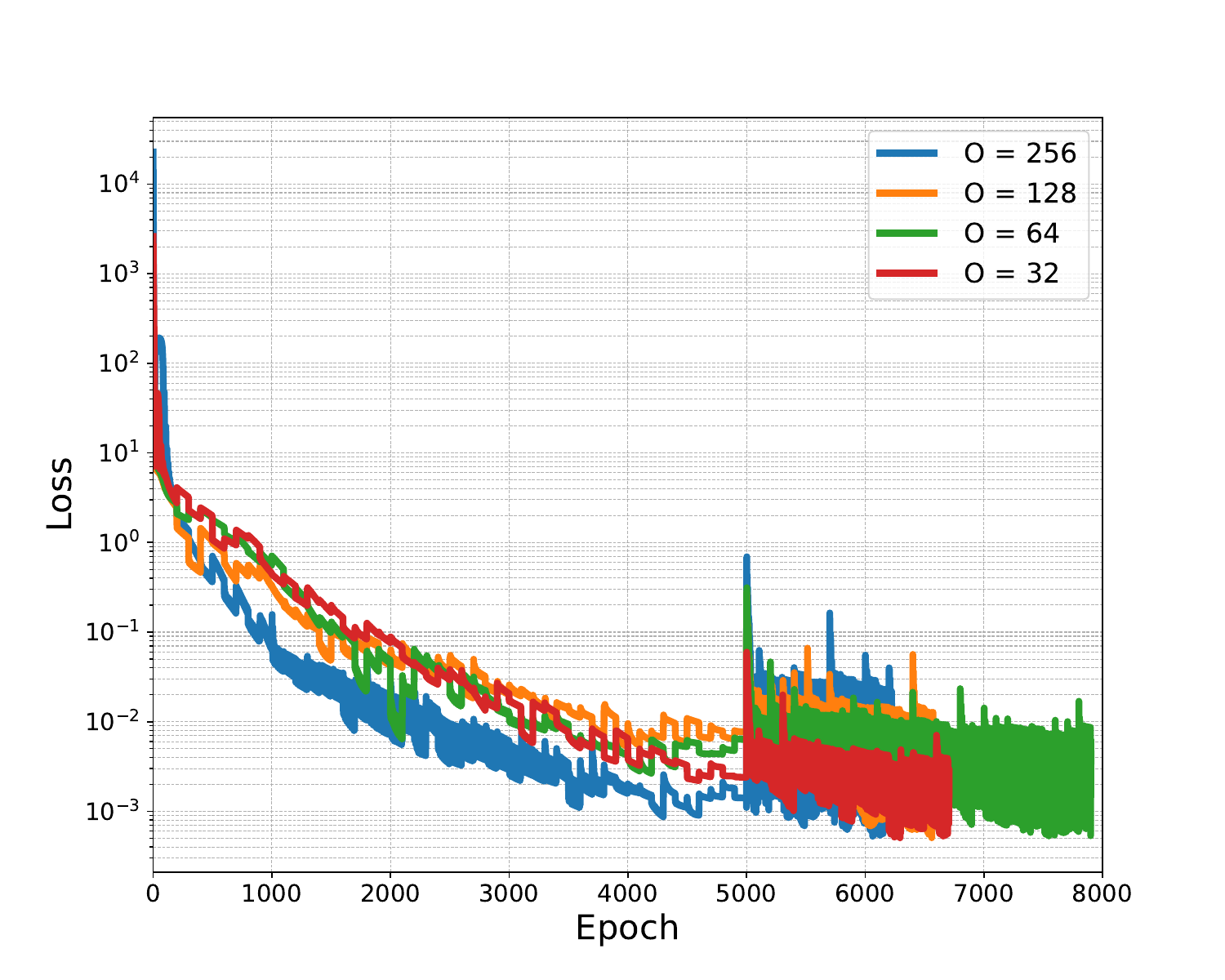}
        \caption{Evolution of the loss function for different number of neurons in a model with $H=6$ hidden layers.}
        \label{fig:aether_loss_neurons}
    \end{subfigure}
    \caption{Loss function evolution with the number of epochs when solving the Lorentz violating gravity equation \eqref{eq:aether} for spin-0  propagation speed squared $s_{0}^{2}=1.5$, and target loss value $\mathcal{T}_{\mathcal{L}}=5\times 10^{-4}$.}
    \label{fig:aetherloss}
\end{figure}

With these choices, we can solve equation \eqref{eq:aether} for different values of the spin-0 propagation speed squared $s_{0}^{2}$. The hyperparameters used to find these solutions with PINNs are summarized in table \ref{tab:aether}. For this equation, we set a target loss value ${\cal T}_{\cal L}=5\times 10^{-4}$ and stop the training once the total loss function reaches ${\cal L}_{\rm \ae ther}<{\cal T}_{\cal L}$. The solutions obtained via the PINNs approach are compared to those obtained by traditional methods in Fig.\ref{fig:aetherfield}, where we can observe excellent agreement between the two. Furthermore, Fig.\ref{fig:aethererror} shows that the relative error between these solutions remains consistently below the $1\% $ level across the spatial domain. Naturally, the error between solutions coming from different methods is inherently tied to the set target loss ${\cal T}_{\cal L}$, and how low can this target loss be set depends on the network's hyperparameters and architecture. It is not the aim of this work to find which configuration of parameters and architectures yields the most accurate solutions or the cheapest/fastest to achieve from a computational standpoint. However, we do explore how the loss function changes as a function of the network's architecture. In particular, we explore how the loss evolution changes with the number of hidden layers and the number of neurons in each layer; Fig.\ref{fig:aether_loss_layers} shows the loss evolution for different number of hidden layers while keeping the number of neurons fixed, while Fig.\ref{fig:aether_loss_neurons} shows the loss evolution for a fixed number of hidden layers for different number of neurons. For this example, all the networks explored perform similarly well; there seems to be only a slight preference for shallower networks with more neurons. 

\begin{table}
    \centering
    \begin{tabular}{|c||c|c|c|c|c|}
        \hline
        Hyperparameter & $\eta_{\rm in}$ & Adam $\beta$'s & $P$ & $S$ & $N$ \\
        \hline
        Value& $3 \times10^{-3}$ & (0.9,0.9) &  $5\times10^3$ & $10^2$ & 100 \\
        \hline
    \end{tabular}
    \caption{Hyperparameters chosen for solving the Lorentz violating gravity differential equations.}
    \label{tab:aether}
\end{table}

\subsection{Spherical accretion}\label{sec:accre}
Finally, we focus on another relevant example from a physics standpoint. This is the problem of spherically symmetric accretion of a perfect fluid into a Schwarzschild black hole. Spherical accretion was discussed first
at Newtonian level by Bondi~\cite{Bondi:1952ni}, while the relativistic solution is due to Michel~\cite{Michel:1972oeq}. While the Michel solution is idealized -- because of the assumption of spherical symmetry and perfect fluid --, it provides an excellent benchmark for relativistic hydrodynamics codes. 

The equations describing the flow of the fluid follow from the covariant conservation $\nabla_\mu T^{\mu\nu}=0$ of the stress-energy tensor $T^{\mu\nu}=(p+\rho) u^\mu u^\nu+p g^{\mu\nu} $ on the Schwarzchild metric $g_{\mu\nu}$, and from the conservation of the baryonic mass $\nabla_\mu (\rho_0 u^\mu)=0$. Here, $p$, $\rho$ and $\rho_0$
are the pressure,  energy density and baryon density of the fluid,  $u^\mu$ is its four-velocity, and $\nabla$ denotes covariant derivatives compatible with the Schwarzschild metric. To close the system, one has to prescribe an equation of state, which here we choose to be given by the polytropic equation of state $p=\kappa \rho_0^\Gamma$, $\rho=\rho_0+p/(\Gamma-1)$, with $\kappa$ and $\Gamma$ constants.

With these assumptions, the flow is described by the following system of coupled differential equations for $\rho_0(r)$ and $v(r)$, the fluid local velocity as measured by an observer at rest~\cite{Michel:1972oeq,Rezzolla:2013dea},
\begin{align}\label{accr1}
    &(v^2 - c_{s}^2)r\frac{\mathrm{d}\ln\rho_{0} }{\mathrm{d}\ln r} = -2v^2 + \frac{M}{y^2 r} \,\\
    &(v^2 -c_{s}^2)\frac{\mathrm{d}\ln y v}{\mathrm{d}\ln r} = 2c_{s}^2 - \frac{M}{y^2 r} \,,\label{accr2}
\end{align}
where we have set the speed of light and the gravitational constant to unity $G = c=1$, $M$ is the Schwarzschild black hole's mass, $c_s(r)$ is the sound speed at each point, given by
\begin{equation}\label{cs}
   c_s(r) \equiv  \sqrt{\frac{\partial p}{\partial \rho}}=\sqrt{\frac{(\Gamma -1) \Gamma  \kappa  \rho _0^{\Gamma }}{\Gamma  \kappa  \rho _0^{\Gamma }+(\Gamma -1) \rho _0}} ,
\end{equation}
and $y(r) = \gamma \sqrt{1 - 2M/r}$, with $\gamma = 1/\sqrt{1 - v^2}$  the Lorentz factor. 
 
As can be seen, equations~\eqref{accr1}--\eqref{accr2} are singular at the critical radius $r_s$ where $v(r_s)=c_s(r_s)$, i.e. where the flow transitions from subsonic to supersonic, displaying yet another example of singular points that can only be obtained implicitly through the solution to the system. In order for the physical quantities and their derivatives to remain finite at $r=r_s$, the right-hand side of the equations needs to vanish there, which in turn implies a relation between the critical radius and the sound speed, \begin{equation}\label{rs}
    r_s=\frac{M(1 + 3 c_s^2)}{2 c_s^2}\,.
\end{equation}

\begin{figure}[]
    \centering
    \begin{subfigure}[b]{0.45\textwidth}
        \centering
        \includegraphics[width=\textwidth]{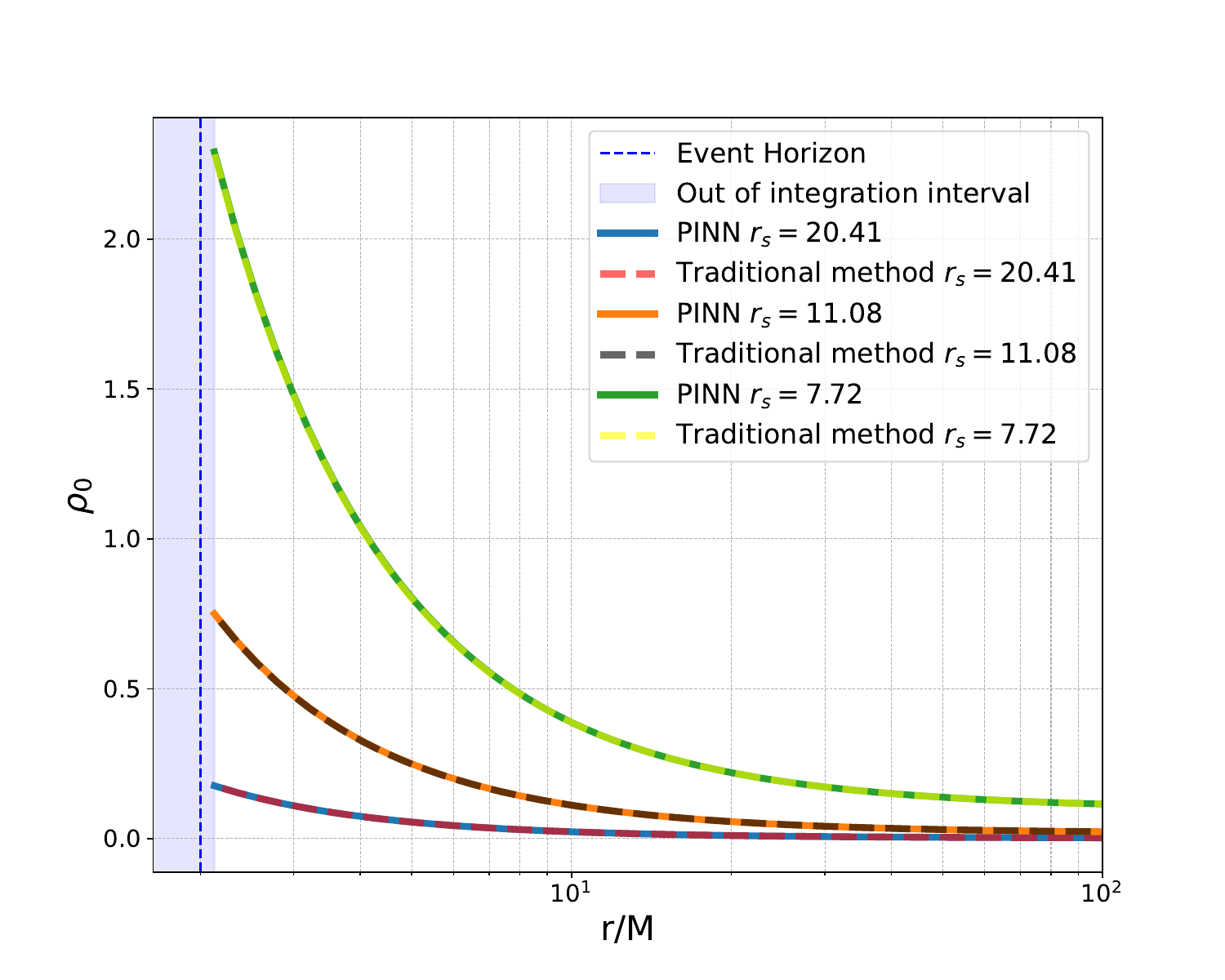}
        \caption{Solutions for the baryonic mass density.}
    \end{subfigure}
    \hfill
    \begin{subfigure}[b]{0.45\textwidth}
        \centering
        \includegraphics[width=\textwidth]{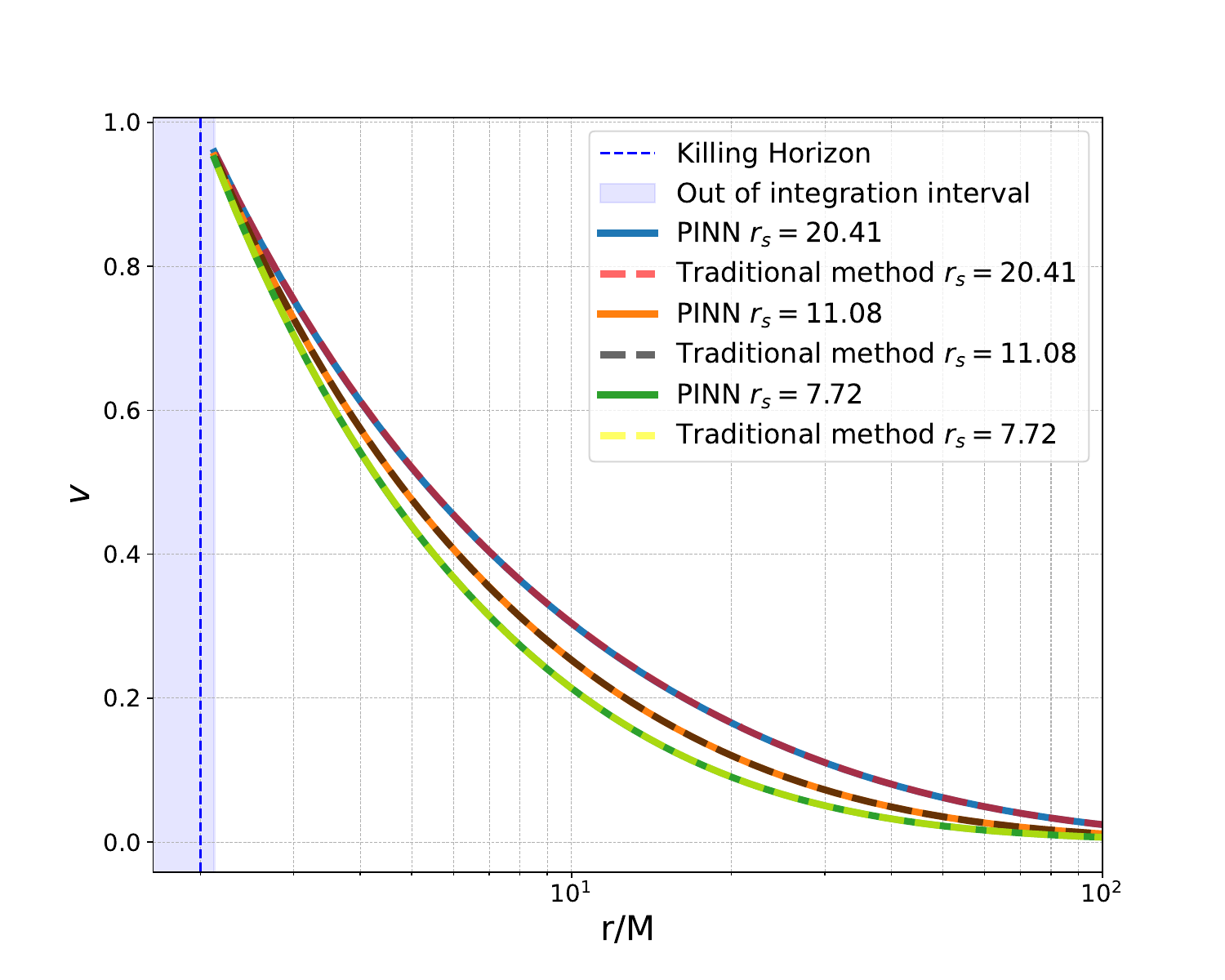}
        \caption{Solutions for the fluid's local velocity .}
        \label{fig:bondivel}
    \end{subfigure}
    \caption{Results for the spherical accretion equations \eqref{cs} for different values of the baryonic density at the critical radius (or equivalently, different asymptotic values), obtained using PINNs, compared to traditional methods. All PINN solutions were obtained with a set of two identical NNs with $H=3$ hidden layers and $O=256$ neurons in each layer. The vertical dashed line in blue denotes the position of the event horizon, and the shaded blue region corresponds to values excluded from the integration regime, which, in this case, corresponds mostly to the interior of the event horizon.}
    \label{fig:bondiv}
\end{figure}

Solving this system with traditional methods can be rather tricky, unless one recognizes that equations~\ref{accr1}--\ref{accr2} can actually be recast as two conservation equations, namely the continuity equation for the baryonic mass and the relativistic Bernoulli equation, so that~\cite{Rezzolla:2013dea}
\begin{equation}\label{cons}
    \rho_0 y v r^2=\mbox{const}\,,\qquad \frac{y(p+\rho)}{\rho_0}=\mbox{const}\,.
\end{equation}
One can then prescribe the baryonic mass density at the critical radius $\rho_0(r_s)$, calculate $c_s(r_s)$ via~\eqref{cs}, and finally $r_s$ by using~\eqref{rs}. The knowledge of these quantities at the critical radius allows in turn for calculating the numerical value of the constants on the right-hand side of~\eqref{cons}. One can then solve~\eqref{cons}
as an algebraic system (e.g. by a Newton-Raphson method) to obtain $v(r)$ and $\rho_0(r)$ at all radii. The solution is shown in figure~\ref{fig:bondiv}, where one can see that the velocity goes to one (i.e. to the speed of light) at the event horizon $r=2M$, as expected since $v(r)$ is the physical velocity measured by a static observer.

Instead, we can use PINNs to solve this problem without recasting it as a set of conservation equations and without the need for a root-finding algorithm. The only requirement is to provide a boundary condition on the value of the baryonic mass density at some arbitrary large radius $\rho_{0}(r_{b})$. We thus implement this condition at $r=r_{b}$ by a parametrization of the solution for the baryonic mass density in terms of the NN and the boundary condition as follows:
\begin{equation}
\rho_{0}^{\theta} =  \rho_{0}(r_{b}) + \frac{r_{b} - r}{r}e^{-n_{\rho}^{\theta}(r)},    
\end{equation}
while the parametrization for the fluid's local velocity is simply,
\begin{equation}
v^{\Theta} =  e^{-n_{v}^{\Theta}(r)},
\end{equation}
where we have again chosen an exponential representation in order to ensure the definite positive character of the solutions.

Moving forward, this time we are dealing with a multi-objective problem, where we use two identical feed-forward NNs to describe both variables, minimizing simultaneously the loss coming from each equation,
\begin{align}
    &{\cal L}_{\rm \rho}= \frac{1}{N}\sum_{a=1}^N\left(  \left( v^{\Theta 2} -c_{s}^2\right)\hat{r}_{a} \left.\frac{d \ln \rho_{0}^{\theta}}{d \ln r}\right|_{\hat r_a} + 2v^{\Theta 2} - \frac{M}{y^{2} \hat{r}_{a}}\right)^2, \\
    &{\cal L}_{\rm v}= \frac{1}{N}\sum_{a=1}^N\left(  \left( v^{\Theta 2} -c_{s}^2\right) \left.\frac{d \ln y v^{\Theta}}{d \ln r}\right|_{\hat r_a} - 2c_{s}^{2} + \frac{M}{y^{2} \hat{r}_{a}}\right)^2, \\
    &{\cal L}_{\rm accr}= \lambda_{\rho}{\cal L}_{\rm \rho} + \lambda_{v}{\cal L}_{\rm v},
\end{align}
where $\lambda_{\rho}$ and $\lambda_{v}$ are determined by the rule \eqref{eq:softadapt}, and we have labeled the total loss function to minimize as ${\cal L}_{\rm accr}$. Furthermore, for this particular example, it proved useful to train the network by employing domain decomposition. If the total spatial domain is given by $\mathcal{D}=\{ r_{f}, r_{b}\} $, we first train the network in a domain $\mathcal{D}_{i} = \{r_{i},r_{b}\}$, where $r_{i} > r_{f}$, until the loss is below a desired threshold $\mathcal{T}_{\mathcal{L}}$. Subsequently, the training continues in a domain  $\mathcal{D}_{i+1} = \{r_{i+1},r_{b}\}$, where $r_{i} > r_{i+1}  >r_{f}$. This procedure can be iterated until reaching the desired spatial domain $\mathcal{D}$, and alleviates problems in enforcing all the domain to converge simultaneously to the same branch of the solution \cite{kapoor2024transfer}. 

With this approach, we can solve equations \eqref{cs} for different boundary conditions. The hyperparameters used are summarized in table \ref{tab:bondi}. In practice, we set a target loss value ${\cal T}_{\cal L}$ and stop the training once the total loss function reaches ${\cal L}_{\rm accr}<{\cal T}_{\cal L}$. In figure~\ref{fig:bondiv}, we see that the obtained solutions are smooth across the whole domain, even through the critical points, and in good agreement with the traditional solutions.

\begin{table}
    \centering
    \begin{tabular}{|c||c|c|c|c|c|}
        \hline
        Hyperparameter & $\eta_{\rm in}$ & Adam $\beta$'s & $P$ & $S$ & $N$ \\
        \hline
        Value& $3 \times10^{-3}$ & (0.9,0.9) &  $10^3$ & $10^2$ & 500 \\
        \hline
    \end{tabular}
    \caption{Hyperparameters chosen for solving the spherical accretion differential equations.}
    \label{tab:bondi}
\end{table}

\begin{figure}[]
    \centering
    \begin{subfigure}[b]{0.45\textwidth}
        \centering
        \includegraphics[width=\textwidth]{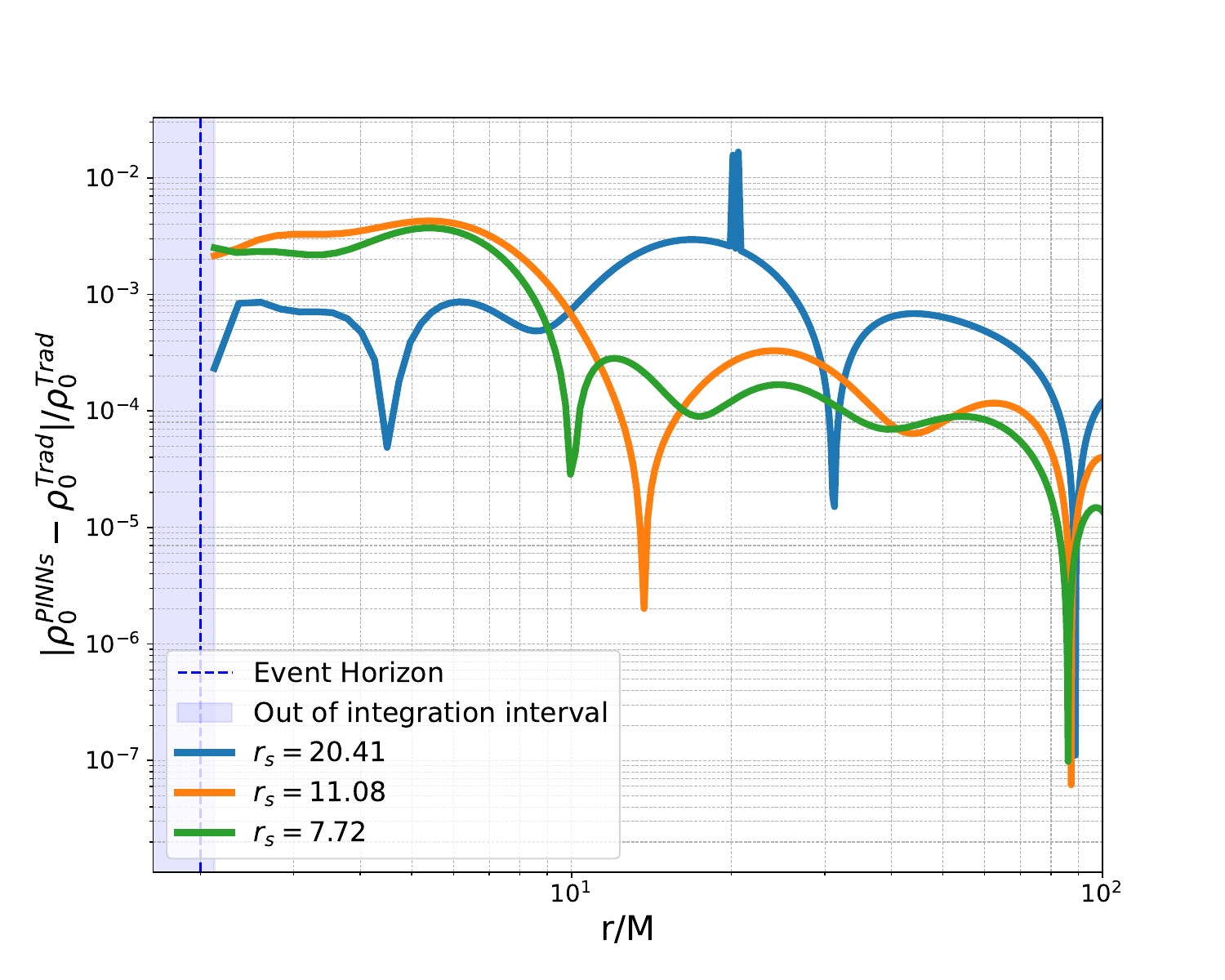}
        \caption{Relative error for the baryonic mass density.}
    \end{subfigure}
    \hfill
    \begin{subfigure}[b]{0.45\textwidth}
        \centering
        \includegraphics[width=\textwidth]{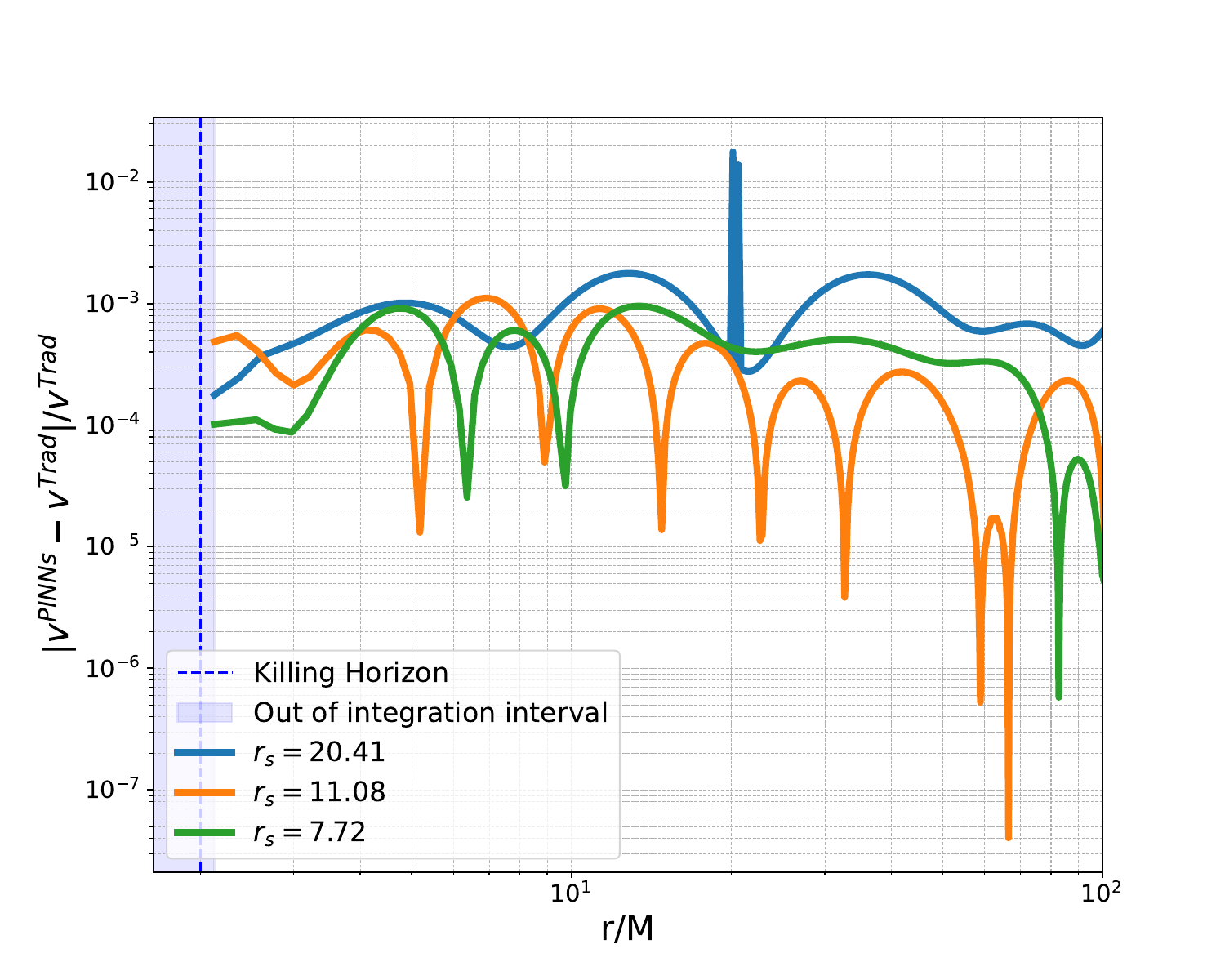}
        \caption{Relative error for the fluid's local velocity .}
    \end{subfigure}
    \caption{Relative error between the PINNs solution and the traditional solutions for different values of the baryonic density at the critical radius (or equivalently different asymptotic values), obtained using PINNs and compared to traditional methods.}
    \label{fig:bondi_err}
\end{figure}

\begin{figure}[]
    \centering
    \begin{subfigure}[b]{0.45\textwidth}
        \centering
        \includegraphics[width=\textwidth]{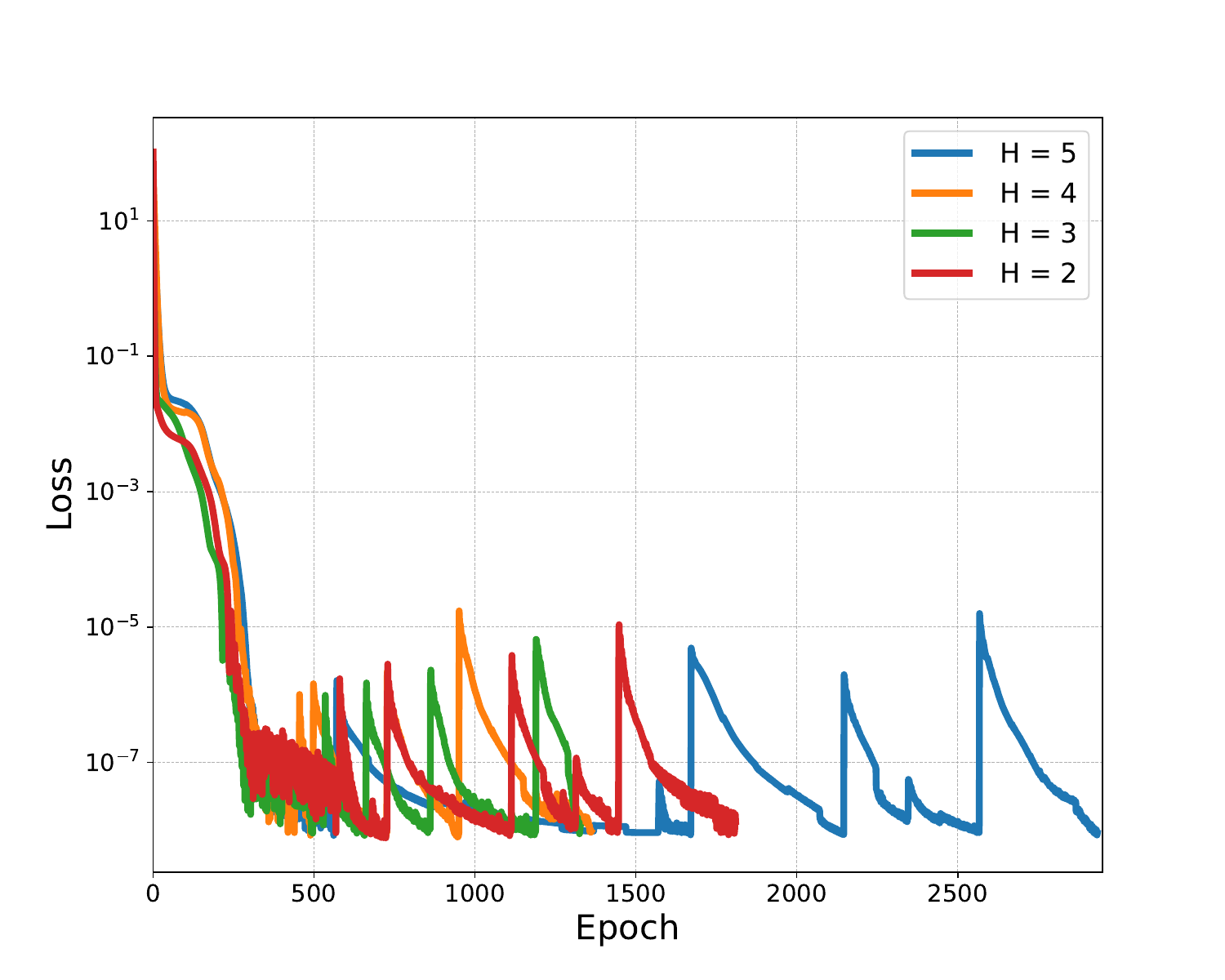}
       \caption{Evolution of the loss function for different number of hidden layers $H$, keeping $O=256$ neurons in each layer.}
    \end{subfigure}
    \hfill
    \begin{subfigure}[b]{0.45\textwidth}
        \centering
        \includegraphics[width=\textwidth]{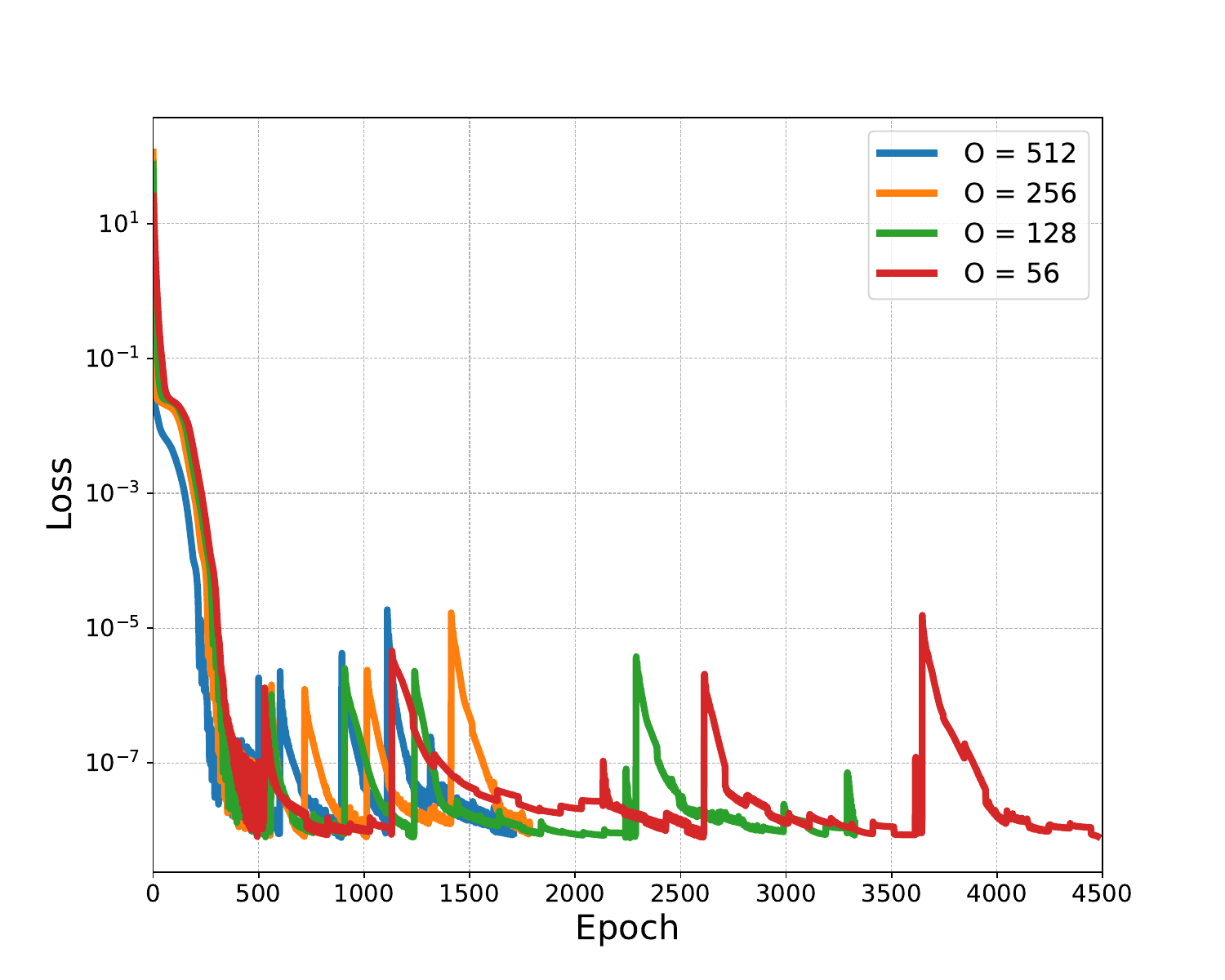}
        \caption{Evolution of the loss function for different number of neurons, in a model with $H=3$ hidden layers.}
    \end{subfigure}
    \caption{Loss function evolution with the number of epochs, when solving the spherical accretion equations \eqref{cs} for a critical radius  $r_{s} = 11.08$, and target loss value $\mathcal{T}_{\mathcal{L}}=5\times 10^{-9}$. }
    \label{fig:bondi_loss}
\end{figure}

Furthermore, figure \ref{fig:bondi_err} shows the relative error between the PINNs solutions and the traditional ones, which remains, for the most part\footnote{The large spike above the percent level is due to the matching error of the solution coming from the traditional method close to the critical radius.}, below the $1\%$ level. The degree of accuracy of our PINNs solutions is ultimately tied to the target threshold for the loss function $\mathcal{T}_{\mathcal{L}}$. How low this threshold can be and how ``quickly'' our solution reaches it depends on the network hyperparameters and architecture. In figure \ref{fig:bondi_loss} we show the loss evolution as a function of the number of epochs for different network architectures, with varying number of hidden layers and neurons. For the hyperparameters chosen, it appears that having a number of layers below $H=5$ is more effective, since we reach the threshold faster, although increasing the number of neurons seems to also help. The sharp spikes of loss increase correspond to the completion of training in a spatial subdomain $\mathcal{D}_{i}$. Indeed, as we enlarge the sub-domain, we expect an initial increase in the loss in the subsequent epochs until convergence is attained again.

\section{Conclusions}\label{sec:conclusions}

Throughout this work, we have investigated whether PINNs can be regarded as an efficient tool to solve ordinary differential equations with singular points, which are common in many physics applications. We have done so by exploring four examples of problems with increasing complexity -- the Legendre equation, the hypergeometric equation, the solution to black hole configurations in Lorentz violating gravity, and the spherical accretion of a perfect fluid on a Schwarzschild background. In principle, the same PINN-based approach can be generalized to the more involved case of {\it partial} differential equations with singular points.\footnote{A trivial example is the wave equation in spherical coordinates, where the center is a regular singular point.}

In all cases, we have found that PINNs are able to converge to the right solutions efficiently. Such success is possible by implementing several techniques that increase the efficiency of training -- adaptive weights, schedule control of the learning rate through cosine annealing, and specific methods to select the collocation points in the integration regime, such as randomization and domain decomposition. Altogether, in the examples explored, these allowed us to reach small values of the loss function in a matter of seconds when running on a NVIDIA GeForce 4070 Super GPU \footnote{The NVIDIA GeForce 4070 Super GPU is far from the raw power found in modern workstation GPUs (for instance, Nvidia H100 series nodes, which are the standard high-efficiency GPUs in modern supercomputers, triple the theoretical performance and VRAM memory of the 4070 Super, while also containing a larger number of more efficient tensor cores).}. Moreover, the training time does not change significantly when the hyperparameters of the networks are varied within reasonable values.

Our results imply that PINNs can be valuable tools to attack numerical problems involving differential equations, either by themselves or in combination with standard methods, which are currently in a much more mature state. One could envision PINNs being used to transverse through singular points, while standard integration methods can be employed to seamlessly extend the solution in the regular region in a more cost-effective manner. Alternatively, one could produce data-points by means of standard algorithms and feed them into a PINN as an unsupervised task in order to increase efficiency of the latter. Many more options might be possible, and we expect PINNs to become standard tools in numerical analysis in the coming future.


\section*{Acknowledgements}
We are grateful to T. Andrade, I. Fridman, P. Solé and R. Srinivasan for discussions. The work of M. H-V. has been supported by the Spanish State Research Agency MCIN/AEI/10.13039/501100011033 and the EU NextGenerationEU/PRTR funds, under grant IJC2020-045126-I; and by the Departament de Recerca i Universitats de la Generalitat de Catalunya, Grant No 2021 SGR 00649. IFAE is partially funded by the CERCA program of the Generalitat de Catalunya. 
E.B. and R.C. acknowledge support from the European Union’s H2020 ERC Consolidator Grant ``GRavity from Astrophysical to Microscopic Scales'' (Grant No. GRAMS-815673), the PRIN 2022 grant ``GUVIRP - Gravity tests in the UltraViolet and InfraRed with Pulsar timing'', and the EU Horizon 2020 Research and Innovation Programme under the Marie Sklodowska-Curie Grant Agreement No. 101007855.
We also acknowledge the use of Google Cloud computing services.

\bibliography{biblio.bib}{}
\end{document}